\documentstyle[epsfig]{mn}

\def\etal{{\rm et al.\thinspace}}
\def\eg{{\rm e.g.\ }}

\def\ie{{\rm i.e.\ }}
\def\cf{{\rm c.f.\ }}

\def\Mdot{\hbox{${\dot M}$} \,}

\def\h50{\hbox{$h_{50}$\,}}

\def\spose#1{\hbox to 0pt{#1\hss}}
\def\ltsimm{\mathrel{\spose{\lower 3pt\hbox{$\sim$}}
	\raise 2.0pt\hbox{$<$}}}
\def\ltsim{$\mathrel{\spose{\lower 3pt\hbox{$\sim$}}
	\raise 2.0pt\hbox{$<$}}$}
\def\gtsimm{\mathrel{\spose{\lower 3pt\hbox{$\sim$}}
	\raise 2.0pt\hbox{$>$}}}
\def\gtsim{$\mathrel{\spose{\lower 3pt\hbox{$\sim$}}
	\raise 2.0pt\hbox{$>$}}$}

\def\fract#1/#2{\leavevmode\kern.1em                   
   \raise.5ex\hbox{\the\scriptfont0 #1}\kern-.1em
   /\kern-.15em\lower.25ex\hbox{\the\scriptfont0 #2}}

\def\cm{{\rm\thinspace cm}}
\def\erg{{\rm\thinspace erg}}

\def\km{{\rm\thinspace km}}

\def\Lsol{\hbox{${\rm\thinspace L_{\odot}}$}}

\def\Msol{\hbox{${\rm\thinspace M_{\odot}}$}}
\def\Rsol{\hbox{${\rm\thinspace R_{\odot}}$}}
\def\pc{{\rm\thinspace pc}}
\def\s{{\rm\thinspace s}}

\def\yr{{\rm\thinspace yr}}
\def\erg{{\rm\thinspace erg}}

\def\cm2{\hbox{${\rm\cm^{2}\,}$}}
\def\pcm2{\hbox{${\rm\cm^{-2}\,}$}}
\def\ergpcm3ps{\hbox{${\rm\erg\cm^{-3}\s^{-1}\,}$}}

\def\kmps{\hbox{${\rm\km\s^{-1}\,}$}}

\def\Lsolppc3{\hbox{${\rm\Lsol\pc^{-3}\,}$}}
\def\Msolppc3{\hbox{${\rm\Msol\pc^{-3}\,}$}}
\def\Msolpyr{\hbox{${\rm\Msol\yr^{-1}\,}$}}

\def\ergps{\hbox{${\rm\erg\s^{-1}\,}$}}

\title[X-ray Analysis of the Eccentric 
O-star Binary Iota~Orionis]
{Coordinated Monitoring of the Eccentric O-star Binary
Iota~Orionis: The X-ray Analysis}
\author[Julian M. Pittard \etal]
  {Julian M. Pittard$^{1,2}$, Ian R. Stevens$^{1}$, Michael F. Corcoran$^{3}$,
         Ken G. Gayley$^{4}$, \cr Sergey V. Marchenko$^{5}$, 
         Gregor Rauw$^{1,6}$\thanks{Charg\'e de Recherches FNRS, Belgium}\\
         $^{1}$School of Physics and Astronomy, University of Birmingham,
                   Edgbaston, Birmingham B15 2TT, UK \\
         $^{2}$School of Physics and Astronomy, University of Leeds,
                   Woodhouse Lane, Leeds LS2 9JT, UK \\
         $^{3}$Universities Space Research Association/Laboratory for 
                   High Energy Astrophysics, GSFC, Greenbelt, MD~20771, USA \\
         $^{4}$University of Iowa, Iowa City, IA~52245, USA \\      
         $^{5}$D\'{e}partment de physique, Universit\'{e} de Montr\'{e}al, 
                   C.P. 6128, Succursale Centre-Ville, Montr\'{e}al,
                   Qu\'{e}bec, H3C~3J7, Canada\\
         $^{6}$Institut d'Astrophysique \& G\'{e}ophysique, Universit\'{e} 
                   de Li\`{e}ge, 5, Avenue de Cointe, B-4000 Li\`{e}ge,
                   Belgium\\
email: jmp@ast.leeds.ac.uk, irs@star.sr.bham.ac.uk,
corcoran@barnegat.gsfc.nasa.gov, kgg@astro.physics.uiowa.edu, \\
sergey@astro.umontreal.ca, rauw@astro.ulg.ac.be}

\date{Accepted .... Received ....; in original form ...}

\pagerange{\pageref{firstpage}--\pageref{lastpage}} \pubyear{2000}

\begin{document}

\maketitle
\label{firstpage}

\begin{abstract}
We analyse two {\it ASCA} observations of the highly eccentric 
O9~III + B1~III binary Iota~Orionis obtained at periastron and apastron. 
Based on the assumption of a strong colliding winds shock between the 
stellar components, we expected to see significant variation in
the X-ray emission between these phases. The observations 
proved otherwise: the X-ray luminosities and spectral distributions were 
remarkably similar. The only noteworthy feature in the
X-ray data was the hint of a proximity effect during periastron
passage. Although this `flare' is of relatively low significance it is
supported by the notable proximity effects seen in the optical
(Marchenko \etal 2000) and the phasing of the X-ray and optical events 
is in very good agreement. However, other interpretations are also possible. 

In view of the degradation of the SIS instrument 
and source contamination in the GIS data we discuss the accuracy of 
these results, and also analyse archival {\it ROSAT} observations.
We investigate why we do not see a clear colliding winds signature.
A simple model shows that the wind attenuation to 
the expected position of the shock apex is negligible throughout 
the orbit, which poses the puzzling question of why the expected 
$1/D$ variation (\ie a factor of 7.5) in the intrinsic luminosity 
is not seen in the data. Two scenarios are proposed: either the 
colliding winds emission is
unexpectedly weak such that intrinsic shocks in the winds dominate the
emission, or, alternatively, that the emission observed {\em is}
colliding winds emission but in a more complex form than we would
naively expect. Complex hydrodynamical models are then analyzed. 
Despite strongly phase-variable emission from the models, {\em both} were
consistent with the observations. We find that if the mass-loss 
rates of the stars are low then intrinsic
wind shocks could dominate the emission. However, when we assume higher
mass-loss rates of the stars, we find that the observed emission
could also be consistent with a purely colliding winds origin. A
summary of the strengths and weaknesses of each interpretation is
presented. To distinguish between the different models X-ray observations
with improved phase coverage will be necessary.
\end{abstract}


\section{Introduction}
\label{sec:intro}
The Orion~OB1 stellar association is one of the brightest and 
richest concentrations of early-type stars in the vicinity of 
the Sun (Warren \& Hesser 1977). It contains large numbers of young O-
and B-type stars, and this, combined with its position well below the galactic
plane ($<b>$ = -16$^{\circ}$) with subsequent low foreground 
absorption, makes it an optimal object to study.
The OB1 association is also famous for an exceptionally dense
concentration of stars known as the Trapezium cluster, located
near $\theta^{1}$~Ori. The stellar density of the Trapezium cluster
has been estimated by several authors in recent years 
and its central region is now
thought to exceed $\sim 4 \times 10^{4}$ stars~${\rm pc^{-3}}$ (McCaughrean \&
Stauffer 1994), making it one of the densest young clusters currently
known.

Iota~Orionis (HR~1889; HD~37043) is a well known highly eccentric ($e =
0.764$) early-type binary system (O9~III + B1~III, $P = 29.13376$~d) 
located in the C4 
subgroup of the Orion OB1 association, approximately 30 arcminutes
South of the Trapezium cluster. In the last decade Iota~Orionis 
has drawn attention for the
possibility of an enhanced, focused wind between the two stars during
the relatively close periastron encounter (Stevens 1988; Gies, Wiggs
\& Bagnuolo 1993; Gies \etal 1996). Although data consistent with this
effect was presented (residual blue-shifted emission in H$\alpha$ 
that accelerates
from about -100~\kmps to -180~\kmps), previous evidence was inconclusive,
and the possibility of a random fluctuation of the primary's wind 
was not ruled out by Gies \etal (1996). These authors also found no 
evidence of any systematic profile deviations that are associated with 
nonradial pulsations. The latest optical monitoring (Marchenko \etal 2000),
again obtained data consistent with a proximity effect.

Iota~Orionis is perhaps even more interesting for the possible 
existence of a strong colliding winds interaction region between the two stars,
which should be most clearly recognized at X-ray wavelengths. As 
demonstrated by Corcoran (1996), perhaps the best direct test for 
the importance of colliding
winds emission is to look for the expected variation with orbital phase.
X-ray data can also be used to extract information on the 
characteristics of this 
region (\eg geometry, size, temperature distribution), which can in turn 
constrain various stellar parameters including mass-loss rates. This 
possibility has already been explored for the Wolf-Rayet (WR) 
binary $\gamma^{2}$~Velorum (Stevens \etal 1996). Furthermore, a direct 
comparison of the recently presented sudden
radiative braking theory of Gayley \etal (1997) with observed X-ray fluxes 
has not yet been made. For this goal, Iota~Orionis has the dual 
advantage of a high X-ray flux (partially due to its relative
proximity, $D \sim 450$~pc) and a highly
eccentric orbit (the varying distance between the stars should act as
a `probe' for the strength of the radiative braking effect). Iota~Orionis
is additionally intriguing for its apparent lack of X-ray variability (see
Section~\ref{sec:prev_x-ray}) whilst colliding wind theory predicts a
strongly varying X-ray flux with orbital phase. 

In this paper we present the analysis and interpretation of two 
{\it ASCA} observations
of Iota~Orionis proposed by the {\it XMEGA} group, as well as a
reanalysis of some archival {\it ROSAT} data. These are the first
detailed {\it ASCA} observations centered on this object, although {\it ASCA}
has previously observed the nearby O-stars $\delta$~Ori and $\lambda$~Ori
(Corcoran \etal 1994) and surveyed the Orion~Nebula (Yamauchi \etal 1996).
{\it ASCA} observations of WR+O binaries (\eg $\gamma^{2}$~Velorum -
Stevens \etal 1996; WR~140 - Koyama \etal 1994) have provided
crucial information of colliding stellar winds in these systems. 
The new {\it ASCA} observations of Iota~Orionis enable us to 
perform a detailed study of the X-ray emission in this system. The
periastron observation covered $\sim 65$~ksec and includes primary 
minimum ($\phi = -0.009$), periastron passage ($\phi = 0.0$), and 
quadrature ($\phi = 0.014$), and allows an unprecedented look at
interacting winds in an eccentric binary system over a range of
different orientations. The apastron observation covered $\sim
25$~ksec and allows us to compare the X-ray properties at the maximum
orbital separation. Thus these two new observations offer the additional
benefit of improving the poor phase sampling of this object.

The new {\it ASCA} observations were coordinated with an extensive
set of ground based optical observations reported in Marchenko \etal
(2000). In this work, the spectra of the components were 
successfully separated, allowing the refinement of the orbital 
elements (including the restriction of the orbital inclination 
to $50^\circ \ltsimm i \ltsimm 70^\circ$) and confirmation of
the rapid apsidal motion. Strong tidal interactions between the components
during periastron passage and phase-locked variability of the
secondary's spectrum were also seen. However, no unambiguous signs
of the bow shock crashing onto the surface of the secondary were found.
These results are extremely relevant to the interpretation of the 
{\it ASCA} X-ray data.
 
This paper is organized as follows: in Section~\ref{sec:prev_x-ray} we
briefly discuss previous X-ray observations of Iota~Orionis; in
Section~\ref{sec:x-ray_analysis} we present our analysis of the new {\it
ASCA} datasets and reexamine archival {\it ROSAT} data. In
Section~\ref{sec:interp} we examine possible interpretations of the
data. Comparisons with other colliding wind
systems are made in Section~\ref{sec:other_cw_systems}, and in 
Section~\ref{sec:conclusions} we summarize and conclude.

\section{Previous X-ray observations of Iota~Orionis}
\label{sec:prev_x-ray}
On account of its high X-ray flux, Iota~Orionis has been observed on
several occasions. X-ray emission from Iota~Orionis was reported 
by Long \& White (1980) in an analysis of {\it Einstein} Imaging 
Proportional Counter (IPC) data, and a count rate of $0.289 \pm 
0.007$~cts/s in the 0.15--4.5~keV band 
was obtained. The spectral resolution ($E/\Delta E$) of the IPC was only 1--2, 
so only the crudest spectral information could be obtained, although
it was observed that the spectrum was soft, peaking below 1~keV.
The authors noted that qualitatively 
good fits could be obtained with a variety of spectral models
leading to an uncertainty of a factor of 2 in 
their quoted luminosity of ${\rm 2.3 \times 10^{32} \ergps}$ (0.15--4.5~keV).

Since this observation the X-ray properties of Iota~Orionis have
been studied on numerous occasions (Snow, Cash \& Grady 1981; Collura
\etal 1989; Chlebowski, Harnden \& Sciortino 1989; Waldron 1991;
Gagn\'{e} \& Caillault 1994; Geier, Wendker \& Wisotzki 1995;
Bergh\"{o}fer \& Schmitt 1995a; Kudritzki \etal 1996; Feldmeier \etal 1997a). 
Due to the different X-ray satellites,
data-analysis techniques, and orbital phases, meaningful
comparisons between the datasets are difficult, and we
simply refer the reader to the above mentioned papers. We note,
however, the results of previous X-ray variability studies of Iota~Orionis,
performed by Snow \etal (1981) and Collura \etal (1989) on {\it Einstein} 
data, and by Bergh\"{o}fer \& Schmitt (1995a) on {\it ROSAT} data. Snow \etal
found that the count rates from 3 IPC observations were twice
as high as the Long \& White observation, leading to the simple interpretation
that the X-ray emission from Iota~Orionis could vary by up to a factor of two
(with a statistical significance of 3$\sigma$), though it was noted
that at the time of the observations there were problems with the instrumental
gain shifts of the IPC. Collura \etal (1989) reexamined the issue of
variability with a more rigorous method and found that they could
only place upper limits of 12\% on the amplitude of variability 
over timescales ranging from 200~s up to the
duration of each observational interval. In contrast to
the findings of Snow \etal (1981) no evidence of long-term variability
between observations was detected. Finally Bergh\"{o}fer \& Schmitt (1995a)
found an essentially flat lightcurve from {\it ROSAT} observations, 
consistent with the previous findings of Collura \etal (1989), although 
again the phase-sampling was rather poor. The lack of dramatic variability 
in these previous studies could be a result of poor phase sampling, and/or the soft
response of the instruments (assuming that X-rays from the wind collision are
harder than those intrinsically emitted from the individual winds). Our {\it ASCA}
observations were specifically designed to address this by i) sampling at phases
where the variation should be maximized; and ii) by leveraging the extended 
bandpass of {\it ASCA} to measure the hard flux.

Table~\ref{tab:prev_obs} summarizes the {\it Einstein} IPC observations 
and the rather confusing situation with regard to whether and at what level 
there is any variability (note that it is {\em not} meant to imply that 
there {\em is} any real source variability). It also emphasizes the poor  
phase sampling and serves as a useful highlight as to one of the reasons why 
the latest {\it ASCA} data presented in this paper were obtained.
In column 3 we list the orbital
phase at the time of observation, calculated from the ephemeris 

\begin{displaymath}
HJD\;(periastron) = 2451121.658 + 29.13376 E
\end{displaymath}

\noindent derived from the latest optical monitoring of this system
(Marchenko \etal 2000). As a result of the large number of orbits that
have elapsed the listed phases are only approximate and do not account for the 
apsidal motion detected in Iota Orionis (Stickland \etal 1987,
Marchenko \etal 2000). We also 
draw attention to the different count rates obtained by different 
authors when analysing the same datasets, and the relatively constant 
count rates obtained from different datasets by the same authors.

\begin{table}
\caption{{\it Einstein} IPC observations of Iota~Orionis.}
\label{tab:prev_obs}
\begin{center}
\begin{tabular}{lllccc}
\hline
Sequence & Date of  & Phase & \multicolumn{3}{c|}{IPC counts/s} \\ 
number   & obs.         & $\phi$     & (LW80) & (SCG81) & (C89)  \\
\hline
3842  & 23/9/79 & 0.34 & 0.289 &       &  \\
11263 & 03/3/80 & 0.90 &       & 0.609 &  \\
5095  & 05/3/80 & 0.97 &       & 0.585 & 0.417 \\
5096  & 23/3/80 & 0.59 &       & 0.607 & 0.413 \\
10413 & 16/2/81 & 0.92 &       &       & 0.367 \\
\hline
\end{tabular}
\end{center}
{\bf Notes.} 
The count rates listed were obtained from the following
papers: LW80 - Long \& White (1980); SCG81 - Snow, Cash \&Grady
(1981); C89 - Collura \etal (1989).
\end{table}

A further {\it ROSAT} analysis of Iota~Orionis was published by Geier 
 \etal (1995), who obtained a position-sensitive 
proportional counter (PSPC) observation centered on the Trapezium
Cluster. This discovered that most of the X-ray emission
from the region originated from discrete sources, in contrast to the
previous {\it Einstein} data where the sources were spatially
unresolved. Nearly all of the 171 X-ray sources were identified with
pre-main sequence stars in subgroups Ic and Id of the Orion OB1
association. The 4 pointings made from March 14-18 1991,
together gave a total effective exposure time of 9685~s. However,
Iota~Orionis was significantly off-axis in all of these pointings. Assuming a Raymond \&
Smith (1977) thermal plasma spectral model, a
single-temperature fit gave $kT = 0.22$~keV and $N_{H} =
10^{20} {\rm cm^{-2}}$, whilst a two-temperature fit with a single
absorption column gave $kT_{1} = 0.12$~keV, $kT_{2} = 0.82$~keV, and
$N_{H} = 3 \times 10^{20} {\rm cm^{-2}}$. 

Two analyses of {\it ROSAT} PSPC spectra which ignored the binary nature
of Iota~Orionis have recently been published by Kudritzki \etal (1996)
and Feldmeier \etal (1997a). The X-rays were assumed to originate 
from cooling zones behind shock fronts. These are a natural development 
of the intrinsically unstable radiative driving of hot-star winds 
(see \eg Owocki, Castor \& Rybicki 1988). Although the authors stated 
that good fits were obtained, no formal indication of 
their goodness was given. Because binarity may significantly affect 
the X-ray emission, we use this as the basis of our interpretation of 
the data. This approach complements these alternative single-star 
interpretations.

\section{The X-ray data analysis}
\label{sec:x-ray_analysis}
Despite previous attempts to characterize the X-ray properties of
Iota~Orionis, our understanding remains poor. For instance, from the
previous {\it Einstein} and {\it ROSAT} data it looks like a typical O-star.
However, on account of its stellar and binary properties, Iota~Orionis should
reveal a clear X-ray signature of colliding stellar winds which one
would expect to show dramatic orbital variability as the separation
between the stars changes. {\it ASCA} should be well suited for
the study of this emission on account of its high spectral resolution
and greater bandpass. In this section we report on the analysis performed 
on the two new phase-constrained {\it ASCA} datasets. This is 
followed by a reanalysis of two archival {\it ROSAT} datasets, and a 
comparison with results obtained from colliding wind models.

\subsection{The {\it ASCA} analysis}
\label{sec:asca_analysis}

\begin{table}
\caption{Effective exposure times and number of counts (background corrected) 
for each of the four instruments onboard {\it ASCA} during the periastron 
and apastron pointings. The percentage of counts in the source region due to 
background varied between 20-25 per cent for the periastron pointing 
and 15-23 per cent for the apastron pointing.
The orbital phase spanning each observation is also listed in column~1.}
\label{tab:my_screen}
\vspace{4mm}
\begin{center}
\begin{tabular}{|c|c|c|c|c|}
\hline
Observation & Inst. & Exp. & Cts in & Ct rate (${\rm s^{-1}}$) \\
(Phase, $\phi$) & & (s) & source & in source \\
                & &     & region & region \\
\hline
Periastron & SIS0 & 65672 & 12543 & 0.191 \\
0.9905-1.068 & SIS1 & 66545 & 9729 & 0.146 \\
           & GIS2 & 72288 & 4301 & 0.060 \\
           & GIS3 & 72336 & 5085 & 0.070 \\
\hline           
Apastron   & SIS0 & 23934 & 4856 & 0.203 \\
1.502 - 1.529 & SIS1 & 23798 & 3225 & 0.136 \\
           & GIS2 & 24496 & 1416 & 0.058 \\
           & GIS3 & 24496 & 1814 & 0.074 \\        
\hline
\end{tabular}
\end{center}
\end{table}

\begin{figure*}
\vspace{80mm}
\includegraphics{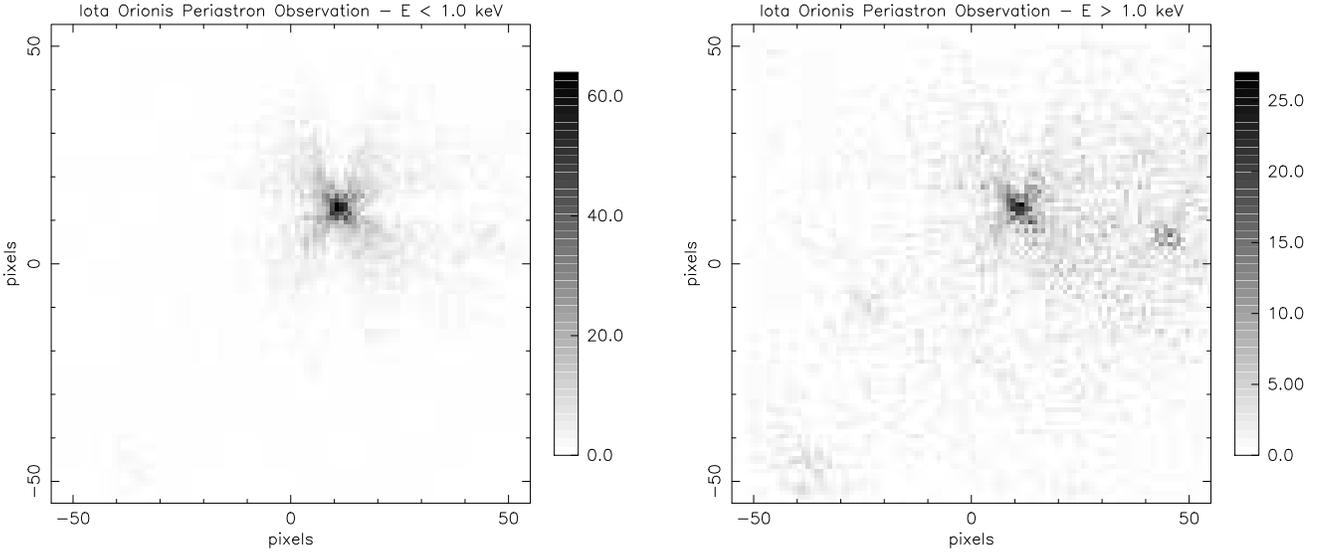}
\vspace{4mm}
\caption{Screened images from the SIS0 periastron data
at energies below 1.0~keV (left) and above 1.0~keV (right). The
serendipitous source is noticeably brighter in the image on the right,
which tells us that its spectrum is relatively harder than
Iota~Orionis. The left image clearly shows the characteristic `Maltese
Cross' of the XRT PSF.}
\label{fig:sis0_peri_E}
\end{figure*}

\subsubsection{Data reduction}
\label{sec:data_red}
Iota~Orionis was twice observed with {\it ASCA} during 1997\footnote{Details 
of the {\it ASCA} satellite may be found in Tanaka, Inoue \& Holt (1994).}. An
observation on 21 September 1997 was timed to coincide with periastron
passage, whilst the 6 October observation was timed to coincide
with apastron passage. The event files from each instrument were 
screened using the FTOOL {\sc
ascascreen}. For the standard SIS analysis the BRIGHT datamode was
processed using medium and high bit rate data. 
Hot and flickering pixels were removed and the
standard screening criteria applied.

The four X-ray telescopes (XRTs) onboard {\it ASCA} each have spatial 
resolutions of $2.9$~arcmin half power diameter. In spite of the broad
point-spread function (PSF), the jittering of the spacecraft can
appear on arcminute scales. To avoid inaccurate flux determinations
and spurious variability in the light curves, the radius of the source
region should normally be no smaller than $3$~arcmin ({\it ASCA} Data
Reduction Guide). For most cases the
recommended region filter radius for bright sources in the SIS is $4$~arcmin.
Because of concerns of contamination from nearby sources,  
our standard analysis adopted a source radius of $3$~arcmin.

The estimation of the X-ray background initially proved somewhat
troublesome. The presence of a nearby source forced a thin annulus,
creating a large variance on the background spectrum, which invariably
still contained contaminating counts. The available blank sky
backgrounds were also found to be unsatisfactory.
After much consideration, the
background was estimated from the entire CCD excluding the
source region and other areas of high count rate. 
This method produced a large number of
background counts, reducing the uncertainty in its spectral shape, and
resulting in a spectral distribution for the source which was better
constrained.

For the GIS analysis, the standard rejection criteria for the 
particle background were used together with the rise time rejection 
procedure. Unlike the SIS, the intrinsic PSF's of the GIS are
not negligible compared to the XRT's and a source region filter
of $4$~arcmin radius was used. The serendipitous
source detected by the SIS instruments was not spatially resolved by
the GIS instruments and hence the source spectra extracted from the
latter contain contaminating counts from this object.

Two methods of background subtraction for the GIS instruments are 
commonly used.
Either one can use blanksky images with the same region filter as used
for the source extraction, or one can choose a source-free area on the
detector at approximately the same off-axis angle as the source. In
our analysis the second method was favoured due to the following reasons:

\begin{enumerate}
\item The cut-off-rigidity (COR) time-dependence can be correctly taken
into account.

\item The blanksky background files were taken during the early stages
of the {\it ASCA} mission, and do not include the secular increase of the GIS
internal background.

\item The blanksky background files were also taken from high Galactic
latitude observations, and hence possible diffuse X-ray emission near
Iota~Orionis cannot be taken into account.
\end{enumerate}

\noindent Despite these reasons, the blanksky method of background
subtraction was also tested and in contrast to
the SIS analysis both methods produced spectral fit results which 
agreed within their uncertainties. However, all subsequent work
including the results reported in this paper used a background
extracted from the same field as the source.

The effective exposure times and count rates for the various
instruments during the periastron and apastron pointings are detailed
in Table~\ref{tab:my_screen}. The numbers in columns 4 and 5 refer only to
the specific source regions extracted.

\begin{figure}
\vspace{85mm}
\includegraphics{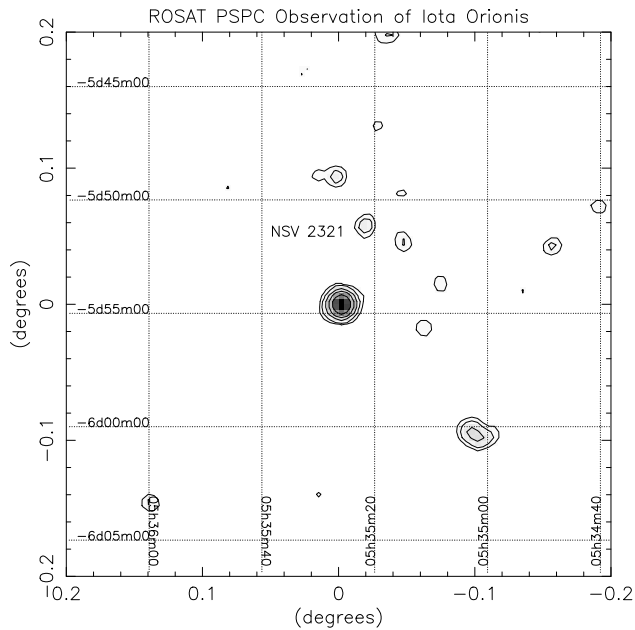}
\vspace{4mm}
\caption{{\it ROSAT} PSPC image of the Iota~Orionis field. This image
number is rf200700n00, and clearly shows the X-ray sources near
Iota~Orionis, which is the brightest source at the centre of the
field. The serendipitous {\it ASCA} source is labelled. Contour levels
are 1, 2, 5, 10, 20 and 50 counts. Epoch 2000.0 coordinates are shown.}
\label{fig:rosat_pspc}
\end{figure}

In Fig.~\ref{fig:sis0_peri_E} we show the SIS0 field from the
periastron observation for two different energy bands. In addition to
Iota~Orionis, a serendipitous source is clearly visible to the right, 
being most
noticeable in the harder image. A {\it ROSAT} PSPC image of
Iota Orionis is shown in Fig.~\ref{fig:rosat_pspc}, and demonstrates more
clearly the crowding of sources in this region of the sky. The
serendipitous source in Fig.~\ref{fig:sis0_peri_E} is tentatively
identified as NSV~2321, an early G-type star (optical coords:
${\rm RA}(2000.0) = 05^{\rm h}~35^{\rm m}~22^{\rm s}$, 
${\rm Dec.}(2000.0) = -05^\circ~54\arcmin~35\arcsec$). As 
already noted, this source complicates the analysis,
of which more details are given in Section~\ref{sec:problems}.

\subsubsection{The X-ray lightcurves}
\label{sec:lightcurves}
\begin{figure*}
\vspace{220mm}
\includegraphics{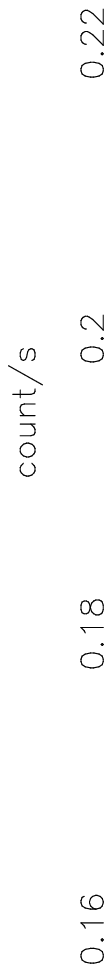}
\vspace{4mm}
\caption{Background subtracted lightcurves from the SIS0 detector for
the periastron (top) and apastron (bottom) observations. They 
are remarkably constant considering the anticipated wind interaction. 
However, an emission `spike' is clearly seen in the periastron lightcurve, 
which occurs just before the point of closest approach of the stars (MJD =
50713.28536). It is not seen in the SIS1 data, though, and we note that 
it occurs just before a period of bad data, and therefore could be entirely
instrumental.}
\label{fig:light_curves}
\end{figure*}

Lightcurves from all 4 instruments were extracted for each of the two 
observations and analysed with the {\sc XRONOS} package. A subset of 
these is shown in Fig.~\ref{fig:light_curves}. Somewhat surprisingly 
the lightcurves were found to be remarkably constant. For the periastron
SIS0 observation, a fit assuming the background-subtracted source was 
constant gave $\chi^{2}_{\nu} = 0.85$. For the apastron observation a 
corresponding analysis gave $\chi^{2}_{\nu} = 0.75$.
If the majority of the X-ray
emission was from the wind collision region, one would expect 
significant variation given the highly eccentric nature of the system. 
There is also little difference between the count rates of the
periastron and apastron observations. 

However, close to periastron passage (corresponding to $2.38 \times 
10^{4}$~s after the start of the observation) there does seem to be a 
small transient spike in both the SIS0 (see Fig.~\ref{fig:light_curves}) 
and GIS2/3 count rates (although nothing appears in the SIS1 lightcurve).
This `flare' is of relatively low significance, 
and could be entirely instrument related
as it occurs just before an interval of bad data. Supporting evidence
for this comes from Moreno \& Koenigsberger (1999) who have recently
investigated the effect of tidal interactions between the stars around 
periastron passage. They found that if the radius of the primary is 
smaller than 15~${\rm \Rsol}$, the enhancement in the mass-loss rate 
at periastron, and therefore the effect on the colliding winds emission,
should not be significant. However, the phasing of the X-ray flare
is almost perfect with a similar event seen in the optical 
(Marchenko \etal 2000). To examine whether the flare possibly occurred
in the background a cross-correlation analysis was performed.
No significant correlation was found between the source 
and background lightcurves, whilst a strong correlation clearly
occurred between the unsubtracted and subtracted source lightcurves. 
We conclude, therefore, that it is unlikely that this `flare' occurs 
in the background data.
 
\subsubsection{The X-ray spectra}
\label{sec:spectra}
\begin{figure}
\vspace{122mm}
\includegraphics{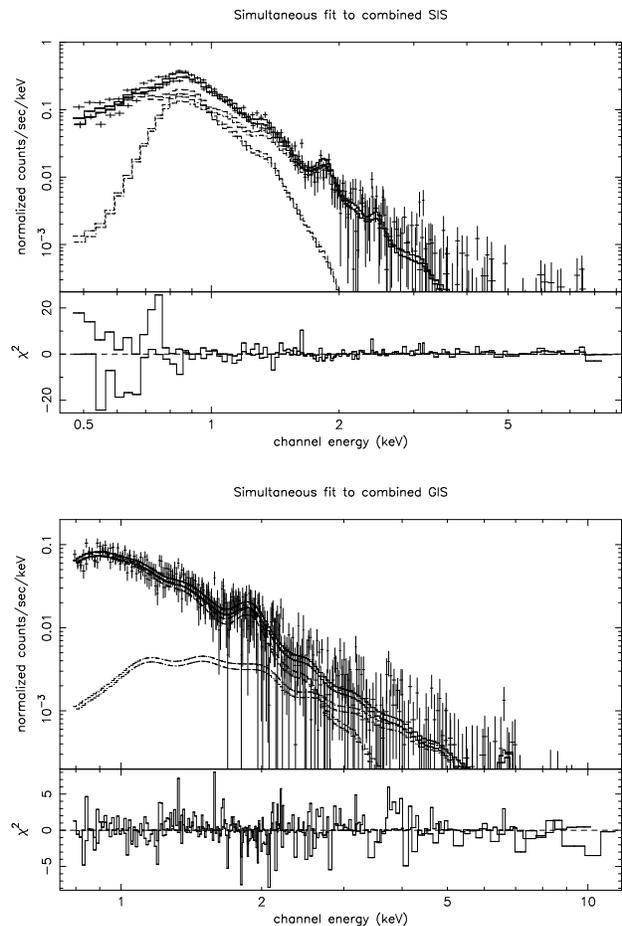}
\vspace{4mm}
\caption{Two-temperature Raymond-Smith spectral model fits to the
combined SIS0 and SIS1 data (upper panel) and the combined GIS2 and GIS3
data (lower panel) from the entire periastron observation. Both 
spectral components and the combined model are shown, together with
the data. A minimum of 10 counts is in each data bin.}
\label{fig:sis0and1_peri_2rs}
\end{figure}

Using the {\sc XSPEC} package, source X-ray spectra were extracted from 
each dataset, re-binned to have a minimum of 10 counts per bin (as 
required for $\chi^{2}$-fitting), and fitted with both single and 
two-temperature Raymond-Smith (RS) spectral models. We emphasize at 
this point that this simple analysis is an effort to 
{\em characterize} i) the overall shape of the
spectrum and its temperature distribution, ii) the amount of variability in
the spectral parameters, and iii) the discrepancies between theoretical
colliding stellar wind spectra and the observed data (see 
Section~\ref{sec:hydro_models}). It is not meant to imply that in the 
two-temperature fits the emission physically occurs from two distinct 
regions at separate temperatures, 
and is simply in keeping with X-ray analysis techniques commonly used today.
The upper panel of Fig.~\ref{fig:sis0and1_peri_2rs} shows a
two-temperature RS model with one absorption component that has been
simultaneously fitted to the SIS0 and SIS1 periastron datasets, whilst
the lower panel shows the corresponding fit to the GIS2 and GIS3
periastron datasets. We include data down to $\sim 0.5$~keV for the SIS
instruments (important for measuring changes in $N_{H}$, although see
Section~\ref{sec:problems}) and to 
$\sim 0.8$~keV for the GIS instruments (where the effective area is 
roughly 10 per cent of the maximum effective area for this instrument, 
thus enabling us to extract maximum information about the emission below 
1.0~keV).

Some line emission is clearly seen, in particular
at $\sim 1.85$~keV and at $\sim 2.30$~keV, of which the former is
most probably from Si{\sc XIII}. Numerous spectral models were fitted
to the data on both an instrument by instrument basis and to various
combinations of the four instruments including all four instruments
together. First we attempted single temperature solar abundance fits. 
Only the fit to the combined GIS2 and GIS3 apastron data was acceptable: 
$N_{H} = 2.71 \pm 0.63 \times 10^{21}$~${\rm cm^{-2}}$, 
$kT = 0.61 \pm 0.06$~keV, $EM = 7.7 \pm 1.7 \times 10^{54}$~${\rm cm^{-3}}$,
$\chi^{2}_{\nu} = 0.92$. Fits using non-solar abundances were
not significantly improved, although the addition of a second independent 
plasma component did result in much better fits and we report some results in
Table~\ref{tab:fit_results}. Immediately obvious is that the fits to the
periastron and apastron SIS data are very similar, both in terms of the
spectral parameters and the resulting luminosities. Clearly there is not the
order of magnitude variation which we expected in the intrinsic luminosity. 
This is also true for the corresponding fits to the GIS data. Thus it appears
at first glance that any variability is small. This has implications for a 
colliding winds interpretation of the data, although at this stage we do not
rule this model out. Other general comments which we can make on the spectral 
fits are:

\begin{enumerate}
\item The SIS datasets return much higher values of $N_{H}$ and much lower
values of the characteristic temperature, $kT$, than the GIS datasets.
 
\item Both the SIS and GIS fits return absorbing columns which are
greater than the estimated interstellar column ($2.0 \times 10^{20}$~$
{\rm cm^{-2}}$; Shull \& Van Steenberg 1985, and Savage \etal 1977) 
arguing for the presence of substantial circumstellar absorption 
consistent with strong stellar winds.

\item Due to the larger variance on the GIS data, the GIS fit results
are statistically acceptable, whereas those to the SIS data are not.

\item If the global abundance is allowed to vary during the fitting
process, the general trend in the subsequent fit results is that the
characteristic temperature increases, the absorbing column decreases,
the normalization increases, and the abundance fits at 0.05--0.2 solar.
The resulting $\chi^{2}_{\nu}$ is basically unchanged. These fit
results bear all the hallmarks of the problems mentioned by Strickland
\& Stevens (1998) and we question their accuracy.

\item The metallicities generally fit closer to solar values if a
two-temperature RS model is used.

\item If a simultaneous SIS and GIS fit is made, the returned
temperatures are similar to those from the SIS dataset alone.
This is perhaps not so surprising, however, given that the variance on the GIS
data is much larger than the SIS, and therefore that the
SIS data has a much larger influence in constraining the fit.
\end{enumerate}

\noindent Two-temperature RS models with separate absorbing columns to
each component were also fitted, as were two temperature spectral models with 
neutral absorption fixed at the ISM value and an additional
photoionized component. However, both failed to significantly improve 
the fitting and we do not comment on these further.

\begin{table*}
\caption{Two-temperature spectral fitting results. Fits made to the combined SIS0 and
SIS1 datasets and the combined GIS2 and GIS3 datasets are shown, as well as that to
an archive {\it ROSAT} dataset (see Table~\ref{tab:rosat_data}). The estimated
ISM column is $2.0 \times 10^{20} {\rm cm^{-2}}$. The emission measure and 
luminosities (0.5 -- 2.5~keV) were calculated
assuming a distance of 450~pc. The emission measure to the soft component
of the rf200700n00 fit could not be constrained and led to unrealistic luminosities. 
Consequently we do not include it here. Many other spectral models were also
examined (see main text) but for conciseness and simplicity we do not
report them here.}
\label{tab:fit_results}
\begin{center}
\begin{tabular}{|c|c|c|c|c|c|c|c|c|}
\hline \hline
Data & $N_{H}$ & $kT_{1}$ & $kT_{2}$ & $EM_{1}$ & $EM_{2}$ & $\chi^{2}_{\nu}$ & $L_x {\rm (int)}$ & $L_x {\rm (abs)}$ \\
 & $(10^{21} {\rm cm^{-2}})$ & (keV) & (keV) & $(10^{54} {\rm cm^{-3}})$ 
 & $(10^{54} {\rm cm^{-3}})$ & (DOF) & $(10^{32} \ergps)$ & $(10^{32} \ergps)$ \\  
\hline
Periastron SIS & $5.07^{+0.28}_{-0.28}$ & $0.14^{+0.01}_{-0.01}$ & $0.61^{+0.02}_{-0.02}$ &
$260^{+61}_{-61}$ & $4.6^{+0.5}_{-0.5}$ & 2.06 (225) & 20.17 & 1.04 \\
Apastron SIS & $3.90^{+0.53}_{-0.53}$ & $0.15^{+0.01}_{-0.01}$ & $0.61^{+0.03}_{-0.03}$ &
$94^{+47}_{-47}$ & $4.1^{+0.7}_{-0.7}$ & 2.06 (143) & 8.99 & 1.00 \\
Periastron GIS & $1.97^{+0.55}_{-0.76}$ & $0.61^{+0.05}_{-0.04}$ & $3.12^{+2.31}_{-0.92}$ &
$6.06^{+0.97}_{-1.21}$ & $0.94^{+0.36}_{-0.34}$ & 1.08 (381) & 2.01 & 1.08 \\
Apastron GIS & $4.73^{+1.43}_{-1.46}$ & $0.25^{+0.10}_{-0.07}$ & $2.15^{+0.70}_{-0.58}$ &
$68^{+266}_{-46}$ & $2.0^{+0.96}_{-0.65}$ & 0.77 (210) & 12.3 & 1.46 \\
rp200700n00 & $0.28^{+0.42}_{-0.15}$ & $0.11^{+0.03}_{-0.05}$ & $0.70^{+0.13}_{-0.14}$ &
$26.7^{+806}_{-14.6}$ & $3.63^{+0.73}_{-0.73}$ & 0.61 (11) & 1.85 & 1.60 \\
\hline
\end{tabular}
\end{center}
\end{table*}

\subsubsection{Problems with the analysis}
\label{sec:problems}

Although the differences between the individual SIS0 and SIS1 datasets, and
between the GIS2 and GIS3 datasets are within the fit uncertainties,
there is a large discrepancy between the fit results made to the SIS
datasets on the one hand and the GIS datasets on the other. Models which fit the GIS
do not fit the SIS, and vice-versa. Whilst we can already state with some
confidence that the X-ray emission is not strongly variable, the above is
obviously a large cause for concern which we should still explore. For 
instance, if we ignore the lack of variability for a moment and assume that a 
colliding winds scenario is the correct model, then the fitted temperatures 
should provide some information on the pre-shock velocities of the 
two winds, and whether these vary as a function of orbital phase.
Hence for this reason we wish to determine which of the SIS or GIS 
temperatures is most accurate.

\begin{figure}
\vspace{65mm}
\includegraphics{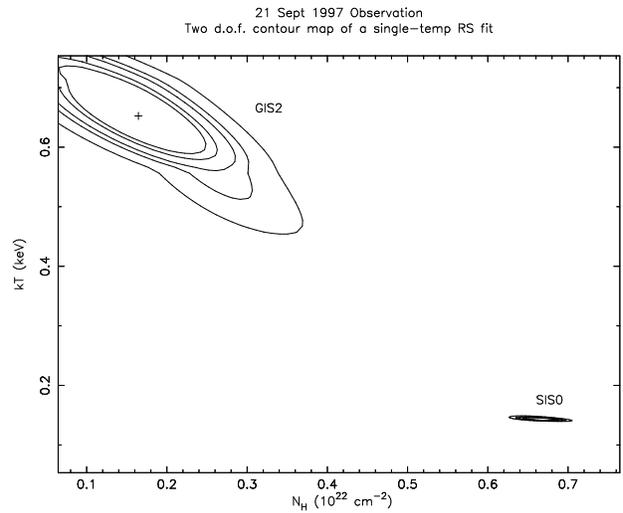}
\vspace{4mm}
\caption{This figure shows the fit-statistic parameter space for a
single-temp RS model fitted separately to the SIS0 and the GIS2
periastron data. The contour levels plotted are the 90.0, 95.4, 99.0,
99.73 and 99.99 per cent confidence regions.}
\label{fig:sisandgis_conf}
\end{figure}

First the fit-statistic parameter space was investigated for false
or unphysical minima. This is shown in Figure~\ref{fig:sisandgis_conf} for a
single-temperature RS model fitted to the SIS0 and GIS2 periastron
data. It is clear that there are no additional minima between the two
shown, and the SIS0 minimum in particular is very compact. However,
at very high confidence levels there is a noticeable extension of the
GIS2 confidence region towards the SIS0 minimum. Despite this the results 
seem to be totally incompatible with each other. Other possible 
explanations for the observed discrepancy are offered below: \\

\noindent \underline{Degradation of the SIS instruments}

\noindent The HEASARC website (http://heasarc.gsfc.nasa.gov/
docs/asca/watchout.html) 
reports that there is clear
evidence for a substantial divergence of the SIS0 and SIS1 detectors since late
1994, with even earlier divergence between the SIS and GIS data. It is
also noted that both the SIS0 and SIS1 efficiencies below 1~keV
have been steadily decreasing over time, which at 0.6~keV can be as much as 20
per cent for data taken in 1998. 
The loss in low-energy efficiency manifests itself as a higher inferred
column density, introducing an additional $N_{H}$ uncertainty of a few
times $10^{20}$~${\rm cm^{-2}}$. This is consistent with our results where 
the SIS detectors return higher values of $N_{H}$ than the GIS (see
Table~\ref{tab:fit_results}), although we note that the difference in
our fits can often be much larger. An increasing divergence of the SIS
and GIS spectra in the energy range below 1~keV has also been noted
by Hwang \etal (1999). As also noted on the HEASARC website a
related problem is that the released blank-sky data cannot
be used for background-subtraction for later SIS data since the
blank-sky data was taken early on in the ASCA mission. This is
also consistent with the problems mentioned in Section~\ref{sec:data_red} in 
obtaining good fits when using this method of background subtraction.
Whilst these problems offer a convincing explanation for the
discrepancies in our fits, we should investigate further before
dismissing the SIS results in favour of the GIS results. \\

\noindent \underline{Contamination of the GIS data}

\noindent The serendipitous source
(hereafter assumed to be NSV~2321) seen in the SIS field is included in 
the GIS spectrum due to the larger PSF of the latter. 
In section~\ref{sec:data_red} it was 
remarked that NSV~2321 was intrinsically 
harder than the Iota~Orionis spectrum (see also
Fig.~\ref{fig:sis0_peri_E}). To examine the effect of NSV~2321 
on the fit results, a $1.06$~arcmin radius circular region 
centred on NSV~2321 was extracted from the
SIS0 periastron data. Although the quality of the spectrum
was poor the fit results confirm that it is a harder source than
Iota~Orionis, which is consistent with the GIS fits having higher
characteristic temperatures than the SIS fits. \\

\noindent \underline{Intrinsic bias}

\begin{figure}
\vspace{205mm}
\includegraphics{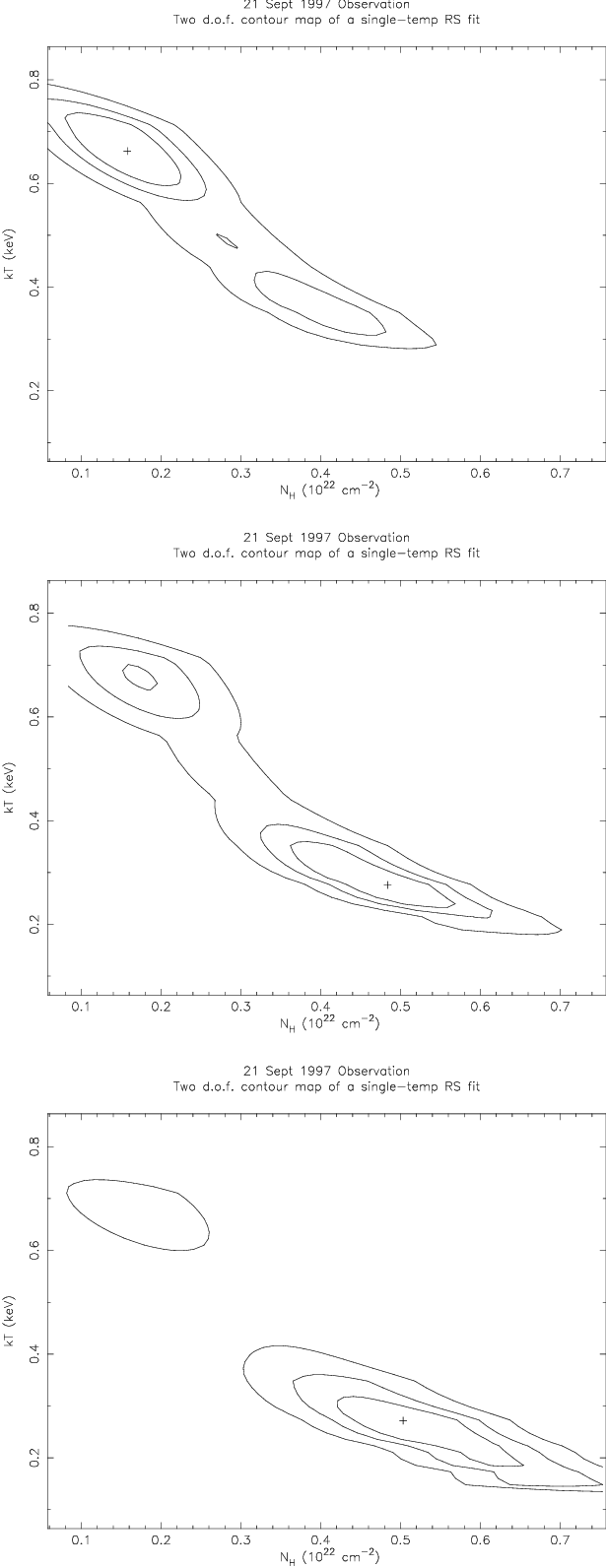}
\vspace{4mm}
\caption{This figure shows the fit-statistic parameter space for a
single-temp RS model fitted to a GIS2 spectrum extracted from a
2~arcmin radius circle centered on Iota~Orionis. The emission above certain
energy thresholds was ignored during the fit process. For the top,
middle and bottom panel these were 2.0, 1.7 and 1.2~keV respectively. 
The contour levels plotted are the 68.3, 90.0 and 99.0\% confidence
regions. The change from one minimum to the other is clearly shown.}
\label{fig:gis2_conf_com}
\end{figure}

\noindent Another possibility for the differences between the SIS 
and GIS results
may lie in the fact that compared with the SIS instruments, the GIS 
instruments are more sensitive at higher X-ray energies. This may result
in fits to the GIS data intrinsically favouring higher characteristic
temperatures. To investigate this we performed fits to a GIS spectrum
whilst ignoring the emission above selectively reduced energy
thresholds. The results are shown in Fig.~\ref{fig:gis2_conf_com}
where a clear bifurcation in the fit results can be seen. As more
high energy X-rays are ignored, the fit jumps from a high temperature
and low column to a low temperature and high column which is much
more compatible with the SIS results. After discovering this behaviour
the SIS0 data was analysed in a similar manner but reversing
the process and selectively ignoring X-rays {\em below} a certain
threshold. Again, a bifurcation develops, this time towards the
original GIS results. This is strong evidence that there is either an
intrinsic bias in the instrument calibrations (either present in the
original calibrations, or a result of the instrument responses
changing with time), or that the fitting process itself is flawed 
\ie that if we fit a
single-temperature spectral model to an inherently multi-temperature
source, the SIS fit will have a lower characteristic temperature than
the GIS fit. Therefore are our fits simply telling us that the source is
multi-temperature? In this case both the SIS and GIS results would be
equally `incorrect'.

\subsection{Re-analysis of archival {\it ROSAT} data}
\label{sec:rosat_analysis}
A search of the {\it ROSAT} public archive for observations of
Iota~Orionis yielded a total of 3 PSPC pointed observations. Two were
back-to-back exposures with Iota~Orionis on axis, of which one used
the Boron filter. The other had Iota~Orionis far off-axis (outside of 
the rigid circle of the window support structure). An advantage of
the {\it ROSAT} observations is that Iota~Orionis is resolved 
from nearby X-ray sources. NSV~2321 is spatially resolved from
Iota~Orionis in two of the three PSPC pointings, and we have
reanalysed both of these. Details of each observation are listed in
Table~\ref{tab:rosat_data}. 

\begin{table}
\caption{Effective exposure times and background subtracted count rate
and number of counts in the source region for the two
{\it ROSAT} PSPC observations examined. The `rf' dataset was 
Boron filtered which blocks X-ray photons with energies between 
0.188-0.28~keV.}
\label{tab:rosat_data}
\vspace{4mm}
\begin{center}
\begin{tabular}{|c|c|c|c|c|}
\hline
Observation & Phase & Exp. & Cts & Ct rate \\
            &       & (s)  &     & (${\rm s^{-1}}$) \\
\hline
rp200700n00 & 0.1697 - 0.1841 & 746 & 1153 & 1.545 \\
rf200700n00 & 0.1841 - 0.1846 & 1305 & 489 & 0.374 \\
\hline
\end{tabular}
\end{center}
\end{table}
                      
\begin{figure}
\vspace{61mm}
\includegraphics{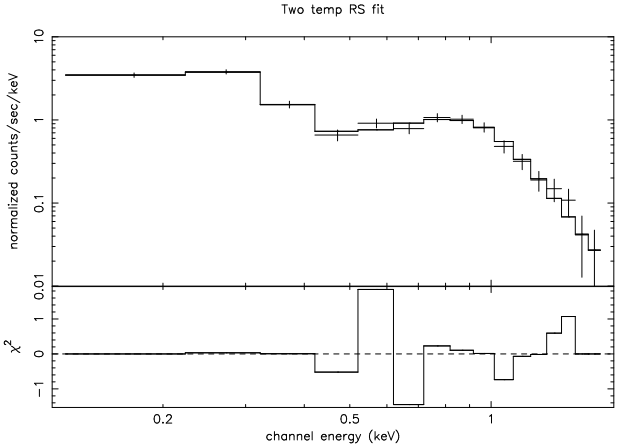}
\vspace{4mm}
\caption{Two-temperature Raymond-Smith spectral model fit to the
rp200700n00 {\it ROSAT} PSPC dataset. The large deviations between the
model and the data at $E \sim 0.5$~keV may be due to uncertainties
in the calibration of the PSPC detector response.}
\label{fig:rosat_spec}
\end{figure}

A single temperature RS model could not reproduce the
observed data for either observation. However, fits with
two characteristic temperatures showed significant improvement and
acceptable values of $\chi^{2}_{\nu}$ were obtained. Unfortunately an
unrealistically soft component was obtained from the Boron filtered dataset
which led to an artificially high luminosity. The fit to the other dataset
is detailed in Fig.~\ref{fig:rosat_spec} and Table~\ref{tab:fit_results}.
We note that if separate columns are fitted to each component the
column to the harder component is much higher than that to the softer
component and the estimated ISM value.

Our results can be compared to the analysis by Geier \etal (1995),
who obtained characteristic temperatures of 0.12 and 0.82~keV from a
two-temperature RS fit made to the rp200151n00 dataset. As
already noted, Iota~Orionis was substantially off-axis in this
pointing and the nearby sources seen in Fig.~\ref{fig:rosat_pspc} were
unresolved. This may account for the slightly higher temperatures that
they obtained.

We can also compare our {\it ROSAT} results to the {\it ASCA} results
obtained in the previous section. In so doing we find distinct
differences between the parameters fitted to the 3 individual
detectors (SIS, GIS and PSPC), although roughly equivalent attenuated 
luminosities. In particular, the {\it ROSAT} spectrum has a 
significantly larger flux at soft energies.

\subsection{The preferred datasets}
\label{sec:preferred_datasets}
This leads us to the crucial question: which dataset do we have most 
confidence in? The negative points of each are:

\begin{itemize}
\item The SIS datasets are invariably affected by the degradation of the CCDs. 

\item The GIS does not have a particularly good low energy response
and does not spatially resolve the close surrounding sources which
contaminate the subsequent analysis.

\item The PSPC has poor spectral resolution compared to the SIS and GIS.
\end{itemize}

\noindent Despite the better spectral resolution of both instruments onboard
{\it ASCA}, it would seem (from the uncertainties mentioned in 
Section~\ref{sec:problems}) that the {\it ROSAT} dataset returns the
most accurate spectral fit parameters. However, the better spectral
capabilities of {\it ASCA} convincingly demonstrates that there is 
no significant 
variation in the luminosity, spectral shape or absorption of the X-ray
emission between the periastron and apastron observations.

\subsection{The X-ray luminosities}
\label{sec:xray_lum}
The attenuated (\ie absorbed) X-ray luminosity in the 0.5--2.5~keV band 
of the rp200700n00 {\it ROSAT} PSPC dataset (fitted by a two-temperature 
RS spectral model with one absorbing column) is 
$\sim 1.6 \times 10^{32} \ergps$.  
As previously mentioned it is often not easy to compare 
values with those estimated in previous papers because of the
different spectral models assumed, the various energy
bands over which the luminosities were integrated, the different
methods of background subtraction employed, and the possible
contamination from nearby sources (\eg NSV~2321). However, as reported in
Section~\ref{sec:prev_x-ray}, Long \& White (1980) estimated an 
attenuated luminosity of $L_{x} = 2.3 \times 10^{32} \ergps$
from a single {\it Einstein} IPC observation.
Chlebowski, Harnden \& Sciortino (1989) obtained $L_{x} = 3.0
\times 10^{32} \ergps$ from seven {\it Einstein} IPC pointings. These values 
are within a factor of 2 of the observed {\it ROSAT} 0.5--2.5~keV 
luminosity from the rp200700n00 dataset ($1.6 \times 10^{32} \ergps$). 
This is not unexpected given the differences in the various analyses. 
The 0.5--10.0~keV luminosities from our analysis of the {\it ASCA} datasets
($ \sim 1.0 \times 10^{32} \ergps$), are also in rough agreement with 
the 0.5--2.5~keV {\it ROSAT} values, albeit slightly reduced
for both the periastron and apastron observations.

\begin{figure}
\vspace{100mm}
\includegraphics{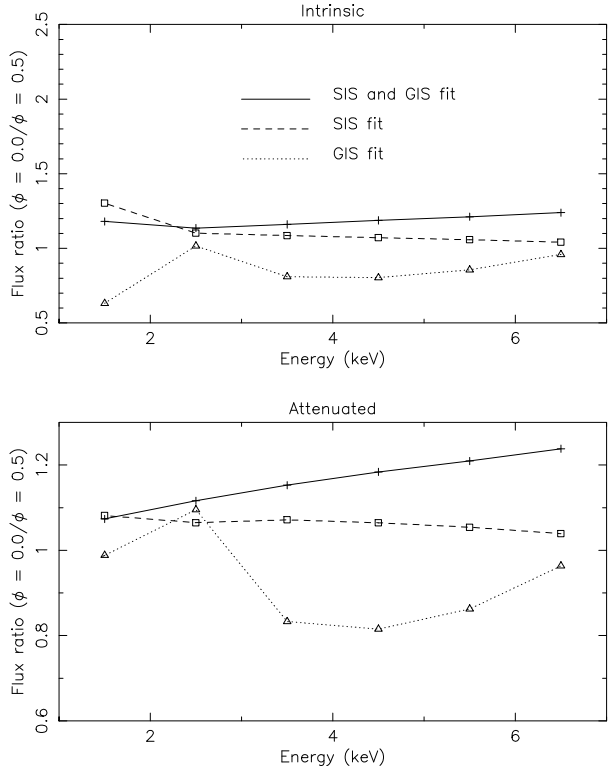}
\vspace{4mm}
\caption{Ratio of the periastron and apastron intrinsic (top) and
attenuated (bottom) fluxes as calculated from a two-temperature RS
spectral model fitted to the {\it ASCA} instruments indicated. The
symbols are plotted at the center of the energy band over which the 
fluxes were integrated (see Section~\ref{sec:xray_lum}).}
\label{fig:lum_ratio}
\end{figure}

Because the {\it ASCA} datasets have much better spectral resolution,
it would be a missed opportunity if we did not take advantage of this.
Thus, whilst again acknowledging that the {\it ASCA} results may
be {\em systematically} incorrect, in Fig.~\ref{fig:lum_ratio} we show the
intrinsic and attenuated flux {\em ratios} of the periastron and apastron
observations. The fluxes are integrated in the energy ranges 1.0--2.0, 
2.0--3.0, 3.0--4.0, 4.0--5.0, 5.0--6.0 and 6.0--7.0~keV although we note
that the counts above 6.0~keV are minimal. Two-temperature
RS spectral model fits were made to the combined SIS0 and SIS1 data, the
combined GIS2 and GIS3 data, and all four datasets together. The top
panel shows the intrinsic flux ratio, whilst the lower panel shows the
attenuated flux ratio. 

Immediately clear from Fig.~\ref{fig:lum_ratio} is the fact that the
intrinsic luminosity from the fit of all four datasets is almost constant
between periastron and apastron, and this is also basically the case
for the attenuated luminosity. If the majority of the X-ray emission
was due to colliding winds, one would expect the X-ray flux at periastron
to be near maximum, since $L_{x}$ is proportional to $1/D$. However,
at energies above $\sim 1$~keV, the intrinsic $L_{x}$ could be severely
reduced around periastron as the shorter distance between the stars
decreases the maximum pre-shock wind velocities. These effects are not
seen.

Assuming that the bolometric luminosity of the Iota~Orionis system is
$L_{bol} = 2.5 \times 10^{5}$~${\rm \Lsol}$ (Stickland \etal 1987), we
obtain log$~L_{x}/L_{bol} = -6.78$ from the {\em observed} {\it ROSAT} 
luminosity. This
compares to a value of $-6.60^{+0.16}_{-0.17}$ deduced from an {\it
Einstein} observation (Chlebowski, Harnden \& Sciortino 1989).
It is in even better agreement with the new results of
Bergh\"{o}fer \etal (1997) who obtain log~$L_{x}/L_{bol} = -6.81$
($\sigma = 0.38$) for stars of luminosity classes III-V and colour
index $(B-V) \ltsimm -0.25$, which are appropriate for Iota~Orionis. 
These results suggest that any emission from a colliding winds shock 
is at a low level ($<$ 50 per cent of the total emission), consistent 
with the lack of variability between the {\it ASCA} pointings. We also 
repeat at this point the comment made by Waldron
\etal (1998) on the validity of the $L{_x}/L_{bol}$ 
relationship: that the canonical ratio 
($L_{x}/L_{bol} sim 10^{-7}$) should only be interpreted as an 
{\em observed} property of X-ray emission from OB stars.

\section{Interpretation}
\label{sec:interp}
In the previous sections we have reported some very puzzling results.
Iota~Orionis was previously thought likely to have a colliding winds
signature which would show significantly different characteristics
between periastron and apastron, and which would act as a test-bed for
the latest colliding wind theories, such as radiative braking.
Although there are concerns over the accuracy of the {\it ASCA} data
and the absolute values of the fit results, it is still possible to
compare relative differences between the periastron and apastron
observations. However, we find that the {\it ASCA} X-ray lightcurves 
and fitted spectra show incredibly little difference between the 
two pointings. 

The extracted spectra also have lower
characteristic temperatures than originally expected, especially for the 
apastron observation where the primary's wind should collide with 
either the secondary's wind or its photosphere at close to
its terminal velocity (see Pittard 1998).

So, the issue is why there is not an obvious colliding winds X-ray
signal in Iota~Orionis. Given the evidence for colliding winds in 
other systems it seems unlikely that there is something fundamentally 
wrong with the colliding winds paradigm. Two interpretations are
therefore possible. One is that the majority of the observed emission is
from intrinsic shocks in the winds of one or both stars (\cf Feldmeier
\etal 1997a and references therein) which dominates a weaker
colliding winds signal. In this scenario the variation of the observed
emission with orbital separation would be expected to be small given the
large emitting volume and the lack of substantial intrinsic variability. 
Alternatively the emission could be from a colliding winds origin but 
in a more complex form than one would naively expect.
In the following subsections we focus our attention on each of these in
turn.

\subsection{The circumstellar absorption}
\label{sec:circum_abs}
Before discussing the merits and faults of each of these scenarios, it
is important to have a good understanding of the level and variability
of the circumstellar absorption in the system.
In Feldmeier \etal (1997a) a model for the wind attenuation of the primary
star was derived from modelling of the UV/optical spectrum. They concluded 
that the wind of the primary O-star is almost transparent to X-rays. 
If the attenuation is indeed negligible then (since the intrinsic 
X-ray colliding winds luminosity is expected to change as $1/D$) we would 
expect to see variability in the {\it ASCA} data. The fact that we do not has
led us to re-consider the attenuation in the system from a colliding winds
perspective, and we have calculated the absorbing column to the apex of the
wind collision region as a function of orbital phase. This 
gives us an idea of the transparency of the wind along the line of
sight to the shock apex. The basic assumptions involved in the
construction of this model were:

\begin{itemize}
\item The winds of both stars are spherically symmetric and are 
characterized by the velocity law

\begin{equation}
\label{eq:beta_vel}
v(r) = v_{\infty} \left(1.0 - \frac{R_{*}}{r}\right)^{\beta}
\end{equation}

\noindent where $\beta = 0.8$.

\item The wind collision shock apex is determined solely by a ram-pressure
momentum balance. No radiation effects such as radiative inhibition or
braking are included. 

\item Where no ram-pressure balance exists the shock collapses onto the
photosphere of the secondary and the apex is therefore located 
where the line of centers intersects the secondary surface.

\item The shock is not skewed by orbital motion.

\item The shock half-opening angle, $\theta$, is determined from the
equation

\begin{equation}
\theta \simeq \left(1 - \frac{\eta^{2/5}}{4}\right)\eta^{1/3}
\end{equation}

\noindent for $10^{-4} \leq \eta \leq 1$ where 

\begin{equation}
\eta = \frac{\Mdot_{2} v_{2}}{\Mdot_{1} v_{1}}.
\end{equation}

\noindent $v_{1}$ and $v_{2}$ are the pre-shock velocities of the primary
and secondary winds at the shock apex. This is basically the equation
presented in Eichler \& Usov (1993) but modified for non-terminal
velocity winds.
\end{itemize}

\noindent The line of sight to Earth from the shock apex is
calculated and the circumstellar absorbing column along it evaluated from 
knowledge of the wind density and the position of the shock cone. The
ISM absorption is not added because it is constant with orbital
phase and we are primarily interested in the variation (and it is also
much smaller at $N_{H} \sim 2.0 \times 10^{20}$~${\rm cm^{-2}}$).

\begin{figure*}
\vspace{70mm}
\includegraphics{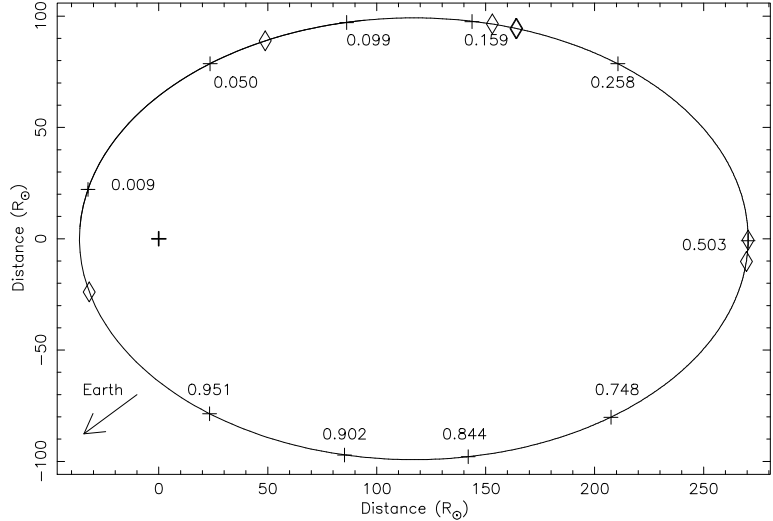}
\vspace{4mm}
\caption{The assumed orbit in the frame of reference of the primary
(marked with a cross at 0,0) for the cw\_1 model. The direction of
Earth is marked with an arrow. At periastron the line of
sight into the system may be through the secondary wind for a very
short period. This is mainly dependent on the inclination angle of 
Iota~Orionis, which is poorly known. The phases, $\phi$, over which 
the {\it ASCA} and {\it ROSAT} exposures were made are delineated by 
diamonds. Also indicated are the phases corresponding to the images 
in Fig.~1 of Pittard (1998) and the 
lightcurve in Fig.~\ref{fig:cw1_att_lum}.}
\label{fig:orbit}
\end{figure*}

\begin{figure}
\vspace{60mm}
\includegraphics{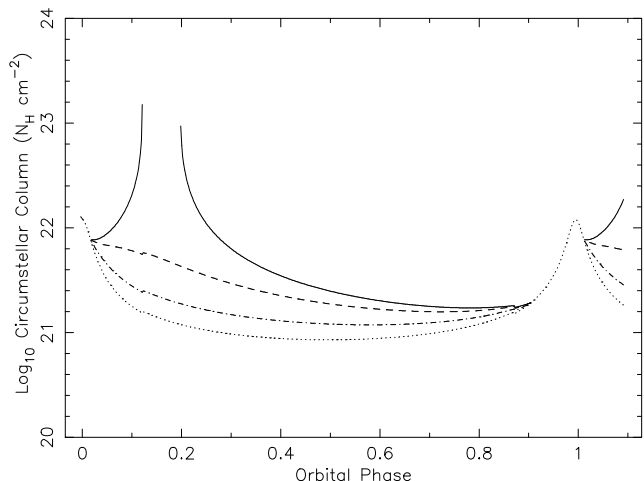}
\vspace{4mm}
\caption{Circumstellar column to the colliding winds shock apex as a 
function of orbital phase for the same stellar parameters as used 
for model cw\_1. The inclination of Iota~Orionis is not well
known so we show results for 4 sets of inclination (solid: $i =
90^{\circ}$; dash: $i = 60^{\circ}$; dot-dash: $i = 30^{\circ}$; dot:
$i = 0^{\circ}$). The latest observational determination puts it at between 50
and $70^{\circ}$ (Marchenko \etal 2000). Gaps in the curves indicate
stellar occultation.}
\label{fig:nh_var}
\end{figure}

The orbit assumed for the calculation is shown in
Fig.~\ref{fig:orbit}. It is based on the cw\_1 model presented in 
Pittard (1998). The direction of Earth is marked, as well as the
position of the secondary star relative to the primary at various
phases. In Fig.~\ref{fig:nh_var} we show the resulting orbital
variability of the circumstellar column with orbital and
stellar parameters appropriate for Iota~Orionis (the values used were 
again for model cw\_1 in Pittard 1998, which for convenience we list in 
Table~\ref{tab:cw_params}) with a range of orbital inclinations. 
For $i = 90^{\circ}$ the numerical results were checked against an 
analytical form. 

\begin{table}
\caption{Parameters used for the hydrodynamical models cw\_1 and cw\_2 -- 
see Pittard (1998) for a discussion of the observational estimates. A
subscript 1 (2) indicates the value for the primary (secondary) star.}
\label{tab:cw_params}
\vspace{4mm}
\begin{center}
\begin{tabular}{|c|c|c|}
\hline
Parameter & Model cw\_1 & Model cw\_2 \\
\hline
$\Mdot_{1} (\Msolpyr)$ & $6.16 \times 10^{-7}$ & $3.06 \times 10^{-7}$ \\
$\Mdot_{2} (\Msolpyr)$ & $1.88 \times 10^{-8}$ & $5.10 \times 10^{-9}$ \\
$v_{\infty 1} (\kmps)$ & $2300$ & $2380$ \\
$v_{\infty 2} (\kmps)$ & $2200$ & $1990$ \\ 
\hline
\end{tabular}
\end{center}
\end{table}

Unlike binaries with circular orbits, the circumstellar column
as a function of phase for an inclination $i = 0^{\circ}$ is {\em not}
constant, being lower at apastron than at periastron. 
In Fig.~\ref{fig:nh_var} this variation is approximately an order of
magnitude. This is a direct result of the density of the primary's wind 
enveloping the shock cone being higher at periastron than at apastron, 
which mostly reflects the combination of the change in the orbital
separation and the pre-shock velocity of the primary.

At higher inclinations ($i \gtsimm 30^{\circ}$) the shock apex is
occulted by the secondary star just before periastron. This is shown
by a gap in the curves in Fig.~\ref{fig:nh_var}. At $i = 90^{\circ}$ 
the primary star also occults the shock apex. On account of the large 
volume of any colliding winds X-ray emission these occultations are 
not expected to have any observational X-ray signatures. At 
inclinations $i \gtsimm 30^{\circ}$ the maximum value of the 
circumstellar column becomes more asymmetric, and shifts 
in phase to where the shock apex is `behind' the
primary (\ie $\phi \sim 0.15$ - see Fig.~\ref{fig:orbit}). The orbital
inclination has just been recalculated to lie in the range $i =
50-70^{\circ}$ (\cf Marchenko \etal 2000). 

Comparing these results with the fitted columns in Table~\ref{tab:fit_results}
is difficult, although they are of the correct order of magnitude.
Assuming an ionized wind temperature of $\sim 10^{4}$~K, the $\tau = 1$ 
optical depth surface occurs at $N_{H} \sim 2 \times 10^{21}$~
${\rm cm^{-2}}$ and $N_{H} \sim 6 \times 10^{21}$~${\rm cm^{-2}}$ for 
0.5 and 1.0~keV photons respectively (\cf for example, Krolick 
\& Kallman 1984). This difference is mostly due to the Oxygen K-edge.
Thus as Fig.~\ref{fig:nh_var} shows, the wind is indeed largely 
transparent at both of these energies, and for realistic values of the 
mass-loss rates. Only at phases near $\phi = 0.0$, where 
$N_{H} \sim 10^{22}$~${\rm cm^{-2}}$, does the absorption begin to
become appreciable (this is one of the causes of the X-ray minima in
Figs.~\ref{fig:peri_lum_drop} and~\ref{fig:cw1_att_lum}).
We therefore face two hard questions. Firstly, given the near 
transparency of the wind throughout most of the orbit, why don't we see 
the intrinsic variation in the colliding winds X-ray luminosity?
Secondly, why do we not see enhanced absorption in our periastron
observation? In the hope of providing satisfactory 
answers to these questions we have calculated the expected colliding winds 
X-ray emission from a complicated hydrodynamical model, as detailed 
in the following section.

\subsection{The cw\_1 and cw\_2 models}
\label{sec:hydro_models}
In this section we investigate in greater depth whether a
colliding winds model {\em can} be consistent with the lack of variability
in the characteristic temperature and count rate of the {\it ASCA} data 
and the unremarkable X-ray luminosity of this system. In particular, we have
two questions: i) is the X-ray luminosity from the wind collision low enough 
that the expected phase variability is lost in the intrinsic background; ii) 
are radiative braking effects stronger than anticipated and can they provide
an explanation for both the low luminosity and the constant X-ray temperatures.

Hydrodynamical models of the wind collision in Iota~Orionis with 
$\eta = 31$ and 72 (with terminal velocity values input as $v_{1}$ and
$v_{2}$) were presented in Pittard (1998). These simulations included the 
realistic driving of the winds by the radiation field of each star by
using the line-force
approximations of Castor, Abbott \& Klein (1975). This approach allows the
dynamics of the winds and radiation fields to be explored, and a
number of interesting and significant effects have been reported (\eg
Stevens \& Pollock 1994; Gayley, Owocki \& Cranmer 1997, 1999). We
refer the reader to these papers for a fuller discussion.

In Pittard (1998) it was found that the colliding winds shock
always collapsed down onto the surface of the secondary star during
periastron passage. On the other hand, there were large differences in the 
dynamics of
the shock throughout the rest of the orbit. For model cw\_2 the 
shock remained collapsed throughout the entire orbit, whilst for model 
cw\_1 the shock lifted off the surface of the secondary
as the stars approached apastron, before recollapsing as the orbit 
headed back towards periastron. However, due to the necessary
simplifications we cannot be sure that the shock actually behaves in this 
manner for a system corresponding to the particular parameters used
(we also reiterate the interpretation of the latest optical data by
Marchenko \etal (2000) who suggest that the shock appears to be
detached even at $\phi = 0.0 - 0.1$). Nevertheless, by comparing 
the orbital variation of the X-ray emission from these models
with the (lack of) observed variability by {\it ASCA} we may be able to 
determine if the observed emission is consistent with a colliding
winds origin and if so whether the collision shock is actually 
detached from the secondary star near apastron. We refer
the reader to Pittard \& Stevens (1997) for specific details of the
X-ray calculation from the hydrodynamic models. We note that an
orbital inclination of $i = 47^\circ$ was assumed for these
calculations (\cf Gies \etal 1996), and we do not expect them to change
significantly in the light of the latest optical determination ($i =
50 - 70^\circ$, Marchenko \etal 2000).

\subsection{An intrinsic wind shock interpretation}
\label{sec:int_wind_shk}

\begin{figure}
\vspace{35mm}
\includegraphics{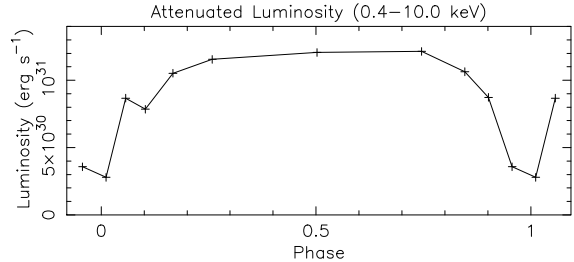}
\vspace{4mm}
\caption{The 0.4--10.0~keV attenuated lightcurve for the colliding wind
model cw\_2. A large drop in the emission around periastron is clearly
seen.}
\label{fig:peri_lum_drop}
\end{figure}

\begin{figure}
\vspace{50mm}
\includegraphics{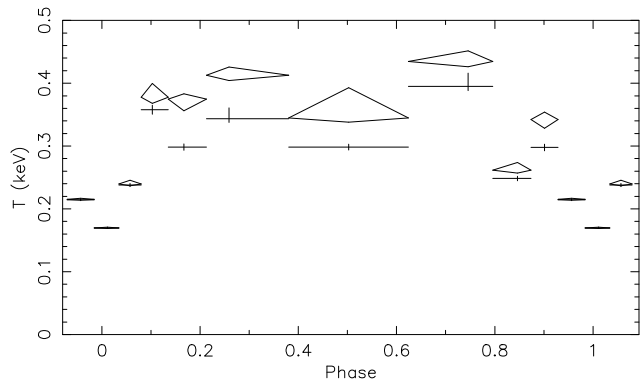}
\vspace{4mm}
\caption{The orbital variation of the characteristic temperature
from a one-temperature RS model fitted to the colliding wind
model cw\_2. Fits assuming solar abundance are marked with crosses, and
those with a global floating abundance are marked with diamonds. 
A minimum in the characteristic temperature around periastron 
is clearly seen, reflecting the reduced pre-shock
velocity of the primary wind in the hydrodynamical model.}
\label{fig:temp_cw2}
\end{figure}

Fig.~\ref{fig:peri_lum_drop} shows the attenuated luminosity from model
cw\_2 as a function of phase. Immediately apparent is its low overall level in 
comparison to the {\it ROSAT} measured luminosity 
($L_{x} \sim 2.0 \times 10^{32} \ergps$). This gives us our first important
insight into this system -- the mass-loss rates of the two stars may be 
low enough with respect to other colliding wind systems that the 
resulting emission is washed out against the
intrinsic background. If this is indeed the case one would not even
expect to notice the severe drop in the colliding winds emission predicted
by the model around periastron. Similarly one would not expect to notice the 
variation in the characteristic temperature of the interaction region shown in
Fig.~\ref{fig:temp_cw2}. We conclude, therefore, that the colliding 
winds X-rays may be so weak that they are washed out by a relatively 
constant component from the intrinsic X-rays from each wind. 
However, we need to be sure that the 
emission measure of this intrinsic component ($EM_{wind}$) is comparable
to the X-ray observations ($EM_{X-ray}$), since if we lower $\Mdot$, we
lower $EM_{wind}$ and could run into the problem where 
$EM_{wind} < EM_{X-ray}$. $EM_{wind}$ can be calculated from:

\begin{equation}
\label{eq:em}
EM_{wind} = \int_{r = R_{min}}^{r = R_{max}} 4 \pi r^{2} \rho^{2}(r) f(r) dr,
\end{equation}

\noindent where $f(r)$ accounts both for the volume filling factor and the 
differences in density of the shocked material relative to the ambient wind
density $\rho(r)$ (which assumes a smooth wind, given by $\rho \equiv 
\Mdot/4 \pi r^{2} v$). Previous wind instability simulations (Owocki \etal 
1988; Owocki 1992; Feldmeier 1995) are broadly consistent with a constant 
or slowly decreasing filling factor as a function of radius (although 
details of the wind dynamics are still largely unknown). 
Feldmeier \etal (1997a) further argue that due to the processes of shock
merging and destruction a monotonically decreasing or roughly constant
filling factor is appropriate. Hence, in the following a constant $f$ will
be supposed. For an instability generated shock model, X-ray emission is
only expected once the wind has reached a substantial fraction ($\gtsimm$ 50
per cent) of $v_{\infty}$. We shall therefore take $R_{min} = 1.5 R_{*}$. 
This is also sensible given that the observed X-ray flux doesn't show 
variability - if the emission occurred too close to the star eclipses 
would occur. With this value we also note that the fraction of the shocked 
wind occulted by the stellar disc is small.
Feldmeier \etal (1997b) have also demonstrated that the X-ray emission from
radii greater than $30 R_{*}$ is negligible, so we take this as our upper
bound, $R_{max}$. This is again reinforced by the fact that in 
Table~\ref{tab:fit_results}, $N_{H} > N_{ISM}$. Equn.~\ref{eq:em} 
then becomes $EM_{wind} = 
f \times EM_{smooth}$, where $EM_{smooth}$ is the emission measure of the
smooth wind evaluated between $R_{min}$ and $R_{max}$. 

Derived filling factors for the shock-heated gas are of order 0.1--1.0 per
cent for O-stars (Hillier \etal 1993), but may possibly approach unity 
for near-MS B-stars (Cassinelli \etal 1994). For model cw\_1, 
$EM_{smooth} = 7.6 \times 10^{56}$~${\rm cm^{-3}}$ (assuming the velocity law 
in Equn.~\ref{eq:beta_vel} with $\beta = 0.8$). When compared to the 
values of $EM_{X-ray}$ in Table~\ref{tab:fit_results} we see that we
require $f \sim 0.001 - 0.42$. For model cw\_2, $EM_{smooth} = 3.0 \times 
10^{56}$~${\rm cm^{-3}}$, and we require $f \sim 0.02 - 1.07$. Despite 
the large ranges these are compatible with the observations (given 
that they also include the difference in density between the shocked 
material and a smooth wind), and therefore are consistent with 
an intrinsic wind shock interpretation.

However, a possible problem with this interpretation arises from the 
larger flux seen in the {\it ROSAT} spectrum at soft energies with 
respect to the {\it ASCA} spectra. If we believe that 
the spectral model parameters from the {\it ROSAT} and {\it ASCA} data are 
inconsistent, and that the variation in circumstellar attenuation 
is negligible, then this must be due to a real change in the 
intrinsic flux. This appears somewhat untenable though, given the
overall lack of variability in single O-stars.

\subsection{A colliding winds interpretation}
\label{sec:cw_interp}

\begin{figure}
\vspace{35mm}
\includegraphics{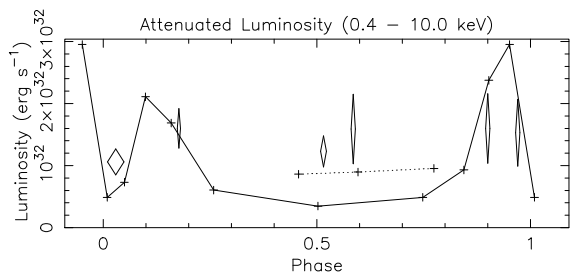}
\vspace{4mm}
\caption{The 0.4--10.0~keV attenuated lightcurve for the colliding wind
model cw\_1. The small crosses (joined by the solid line) were 
calculated from the frames in Fig.~1 of Pittard (1998) at the phases 
indicated on Fig.~\ref{fig:orbit}. 
A large drop in the emission around periastron is clearly
seen.  However, when the shock detaches from the secondary 
in this model a much larger grid is required to `capture' all of the 
X-ray emission -- this is shown by the dotted line. Also marked on
this figure are the observed luminosities from the {\it ASCA} and 
{\it ROSAT} pointings. The uncertainty in the count rates is a
minimum of 1.2 per cent for the SIS0 periastron data, rising to a
maximum of 4.3 per cent for the GIS2 apastron data. However, due
to the uncertainties in the spectral shapes we conservatively 
assume a $\pm$ 20 per cent error in their resultant luminosities. 
A very good fit 
is obtained if one retains a healthy scepticism about the exact
size of the periastron minimum in the model, where our assumption
of axisymmetry is least valid. We also mark on
the relative count rates of the three {\it Einstein} observations 
reported by Snow, Cash \& Grady (1981). Because the conversion from
count rate to luminosity is poorly known for these observations we
have assumed uncertainties of $\pm$ 35 per cent and arbitrarily 
adjusted their scaling to best fit our model curve.}
\label{fig:cw1_att_lum}
\end{figure}

In Fig.~\ref{fig:cw1_att_lum} we show the X-ray
lightcurve from model cw\_1. This time the colliding wind 
emission is much stronger (owing to the higher mass-loss rates assumed -- 
see Table~\ref{tab:cw_params}), and is again very variable. At the phases 
corresponding to the {\it ASCA} data
the synthetic lightcurve is at local minima, and consistent with the observed 
luminosity\footnote{We note that the equations of Usov (1992) give 
un-attenuated luminosities roughly an order of magnitude below the results of
our numerical calculations.}. The model lightcurve also predicts that the
observed luminosity at phase $\phi = 0.17$ should be higher than those at
phases $\phi = 0.0$ and 0.5, with the latter at roughly the same 
level\footnote{ 
There is some uncertainty in the true luminosity during the periastron
passage because the assumption of axisymmetry is poor at these phases, and 
additionally the optical data suggests that the wind collision region
does not collapse onto the surface of the secondary.}.
From Table~\ref{tab:fit_results} we indeed find that this is the case, 
with the {\it ROSAT} luminosity matching the predicted lightcurve almost
exactly\footnote{We note that the quoted GIS $L_{x}$~($\phi = 0.5$) is
much higher than the GIS $L_{x}$~($\phi = 0.0$) value. We conclude
that this is due to uncertainties in the spectral fit parameters
because the GIS background subtracted count rates are nearly identical
from Table~\ref{tab:my_screen}.}. One slight worry is that the 
two {\it Einstein} observations at phases $\phi \sim 0.90$ and 0.97 
do not show a significant enhancement in count rate with respect to the
observation at phase $\phi = 0.59$. It would therefore be clearly
useful to obtain new observations at these phases.

\begin{figure}
\vspace{50mm}
\includegraphics{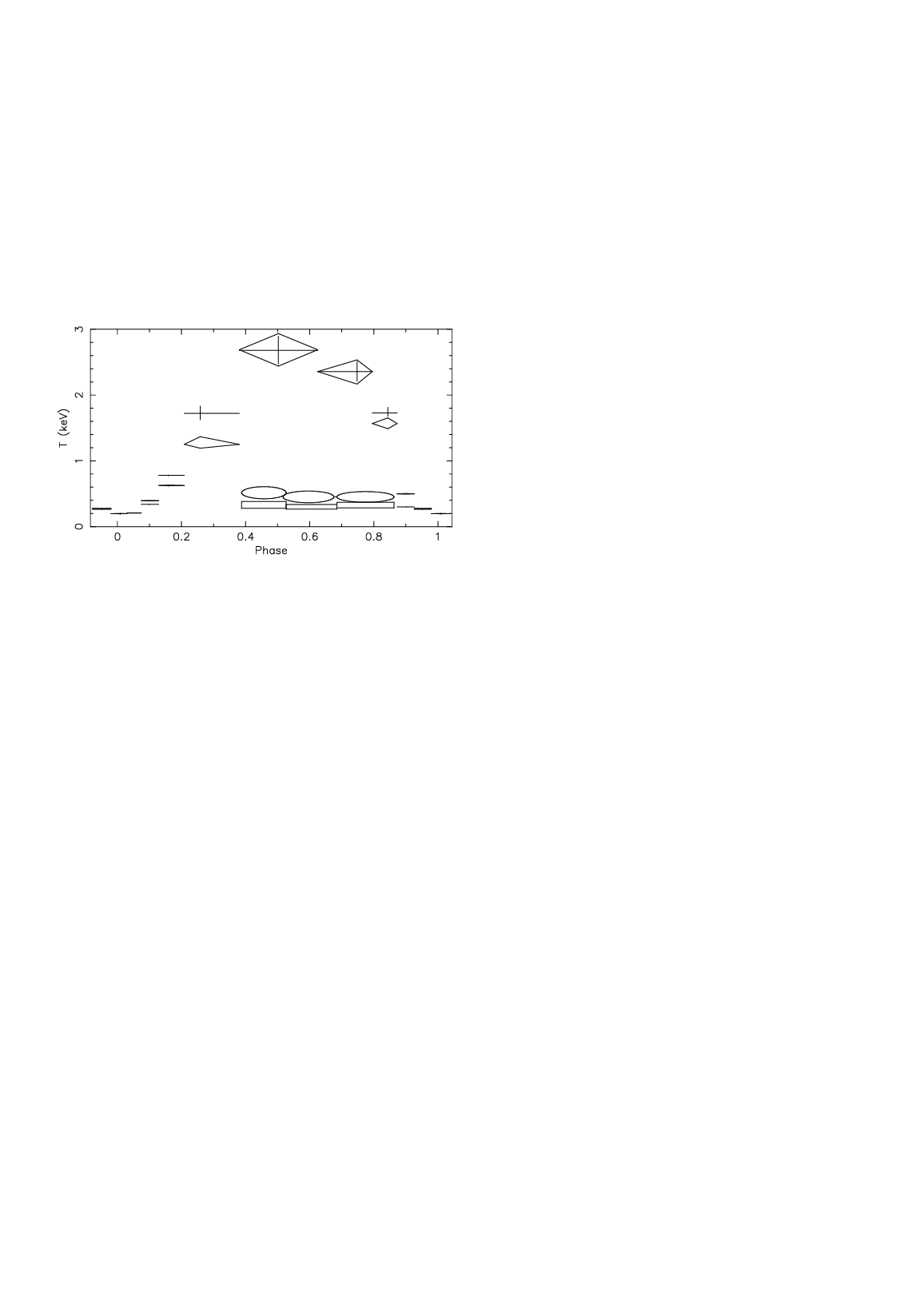}
\vspace{4mm}
\caption{The orbital variation of the characteristic temperature
from a one-temperature RS model fitted to the colliding wind
model cw\_1. Fits assuming solar abundance are marked with crosses, and
those with a global floating abundance are marked with diamonds. 
A minimum in the characteristic temperature around periastron 
is clearly seen, reflecting the reduced pre-shock
velocity of the primary wind in the hydrodynamical model. Note that
unrealistically high values ($> 1$~keV) are obtained over the range
$0.2 \ltsimm \phi \ltsimm 0.9$ due to the grid not capturing all
of the emission. The rectangles and ellipses show the correct 
results of solar and floating abundance fits to a larger 
hydrodynamical grid over this phase range. Thus the characteristic
temperature is within the range 0.2 -- 0.4~keV over the {\em entire}
orbital phase.}
\label{fig:temp_cw1}
\end{figure}

The orbital variation of the characteristic temperature fitted to the
colliding winds region of model cw\_1 is shown in Fig.~\ref{fig:temp_cw1}.
Our best estimate of the temperature varies from $\sim 0.2$~keV at
periastron to $\sim 0.3-0.4$~keV at apastron. This temperature variation
is much less than expected from a simple consideration of the variation
of the pre-shock velocities along the line-of-centres, and reflects
the fact that the fitted temperature is an average over the entire
post-shock region. The temperatures predicted by the model are
comparable with the observed data and we thus conclude that this model 
demonstrates that the observed emission does not allow us to discard
a colliding winds interpretation.

\subsection{Other possibilities/factors}
\label{sec:other_factors}

Another intriguing possibility is that radiative braking in the 
Iota~Orionis system is actually quite efficient (moreso than the 
predictions of both models cw\_1 and cw\_2). This might also explain 
the inference of Marchenko \etal (2000) that the shock is
lifted off the surface of the secondary around periastron. This would require
strong coupling between the B-star continuum and the primary wind, but 
cannot be ruled out by our present understanding.
In the adiabatic limit (which is a good approximation for the wind 
collision in Iota~Orionis at apastron), Stevens \etal (1992) determined that

\begin{equation}
L_{x} \propto \Mdot^{2} v^{-3.2} D^{-1} (1 + \sqrt{\eta})/\eta^{2},
\end{equation}

\noindent where it was assumed that the temperature dependence of the cooling 
curve was $\Lambda \propto T^{-0.6}$. This latter assumption is appropriate 
for post-shock gas in the temperature range $\sim 10^{5} - 10^{7}$~K, 
which corresponds to pre-shock velocities of up to $\sim 1000 \kmps$ for 
solar abundance material. (Note: there is a typographical error in 
the corresponding 
equation of Stevens \& Pollock, 1994). Assuming that $\eta$ is small (\ie 
that the wind of the primary dominates) we then find that to first order, 
$L_{x} \propto v^{-1.2}$ Thus the X-ray luminosity is almost 
inversely proportional to the pre-shock velocity of the primary wind.
A reduction of 10 (50) per cent in the latter leads to an increase in the 
emission of 13 (130) per cent. Conversely, we find from the equations 
in Usov (1992), that $L_{x} \propto v^{-5/2}$ for the shocked primary 
wind and $L_{x} \propto v^{1/2}$ for the shocked secondary 
wind. For model cw\_1 the emission from the latter appears to be 
dominant. Hence, for the same variations in the pre-shock velocity as 
discussed above, the emission is {\em reduced} by 5 and 30 per cent 
respectively. The discrepancy between these two sets of results is 
due to the different assumptions on the form of the cooling curve. 
Usov (1992) assumed cooling dominated by bremsstrahlung 
(\ie $\Lambda \propto T^{1/2}$), which is more appropriate at higher 
temperatures ($T > 10^{7}$~K). Clearly then, the situation is rather 
confusing at present, not least because the {\em actual} position of the 
colliding wind region relative to the two stars is very poorly known, 
as is the relative mass-loss rates of the stars.
It is therefore difficult for us to be any more quantitative without 
performing a rigorous parameter-space study.

It is also possible that the characteristics of the emission may be
altered by other physical processes. In investigating how electron
thermal conduction may affect colliding wind X-ray emission, Myasnikov
\& Zhekov (1998) discovered that pre-heating zones in front of the
shock have the overall effect of increasing the density of the wind
interaction region. The resulting X-ray emission was found to change
markedly, with a large increase in luminosity and a significant
softening of the spectrum. Softer X-rays suffer significantly higher
absorption so the resulting observed emission may have an unremarkable
luminosity. Whether the observed emission would show orbital
variability is not clear at this stage.

Finally, complex mutual interactions between the various physical processes
occurring in colliding winds systems may also significantly alter the resultant 
emission. For instance, Folini \& Walder (2000) suggest that the effects of
thermal conduction and radiative braking may positively reinforce each other.
We are a long way from performing numerical simulations which would 
investigate this.

\section{Comparison with other colliding wind systems}
\label{sec:other_cw_systems} 
Most early-type systems with strong colliding wind signatures are 
WR+OB binaries. These are different from OB+OB systems in a number of ways 
which may explain why it appears that stronger colliding wind signatures 
are obtained from WR+OB systems. First, the high values
of $\Mdot_{WR}$ provide more wind material which can be shocked.
Second, the spectral emissivity for WC wind abundances is greater 
than that for solar abundances at the same
mass-density (\ie mass-loss rate -- see Stevens \etal 1992). However, both of
these points may also act in reverse (high mass-loss rates
also provide more absorption, and the emissivity of WN wind abundances
for a given mass density is below that of solar).
At this point it is therefore instructive to compare the X-ray
characteristics of Iota~Orionis with other early-type binary systems.
The latter can be subdivided into three distinct groups as detailed in
the following subsections.

\subsection{Those with strong colliding winds signatures}
Into this group fall the well-known binaries WR~140 (HD~193793) and
$\gamma^{2}$~Velorum (WR~11, HD~68273). Both these systems show clear evidence 
for colliding stellar winds including phase-variable emission and a hard
spectrum. WR~140 (WC7 + O4-5, $P$=7.94~yr) is famous for its
episodic dust formation during periastron passage (Williams 1990).
Strong X-ray emission from WR~140 was discovered by {\it EXOSAT}
and its progressive extinction with phase by the
WC7 wind has been used to derive the CNO abundances of the WR wind
(Williams \etal 1990). Possible non-thermal X-ray emission has also 
recently been discovered in this system, as witnessed by the 
variability of the Fe~K line (Pollock, Corcoran \& Stevens 1999). 

{\it ROSAT} observations of $\gamma^{2}$~Velorum (WC8 + O9I, $P$=78.5~d) 
were presented by Willis, Schild \& Stevens (1995).
The hard X-ray flux showed a phase repeatable increase 
by a factor of 4 when the system was viewed through the less-dense
wind of the O-star, which was attributed to the addition of a harder
component to the largely unvarying softer emission. More recent 
{\it ASCA} observations (Stevens \etal 1996, and a re-analysis 
by Rauw \etal 2000) confirmed these results. 
A detailed comparison of this data with synthetic spectra generated
from a grid of hydrodynamical colliding wind models provided
evidence that the hard X-ray emission comes directly from
the wind collision. 

\subsection{Those showing some evidence of colliding winds}
The best example of this category is V444~Cyg (WR~139, HD~193576), 
a well-studied eclipsing WN5 + O6 binary with an orbital 
period $P$=4.21~d and well known physical parameters.
Corcoran \etal (1996) confirmed from phase-resolved {\it ROSAT} observations 
that although the luminosity was 1--2 orders of magnitude lower than the 
predictions of Stevens \etal (1992), there was an orbital dependence 
of the flux. This on its own did not completely rule out the O-star 
being the source of the X-rays, though fairly convincing evidence 
for colliding winds remained. First, the high value of the 
characteristic temperature outside of eclipse ($kT \sim
1.7$~keV) was much greater than the typical temperatures of single
O-stars ($kT \sim 0.5$~keV). Second, the variation of the emission
was consistent with a shock location near the O-star surface, as 
would be expected given the higher mass-loss rate of the WR-star and 
previous deductions from ultra-violet {\it IUE} observations 
(Shore \& Brown 1988). Finally,  
the assumption of instantaneous terminal wind velocity 
overestimates the observed X-ray luminosity.
More recent {\it ASCA} (Maeda \etal 1999) and optical (Marchenko \etal 1997) 
observations of this system have reinforced this interpretation. 
However, some puzzling X-ray characteristics remain such as large amplitude 
short time-scale variability which is not easily explained by a 
colliding winds model (Corcoran \etal 1996).

Another system which shows some, but not overwhelming, characteristics 
of colliding winds is 29~UW~Canis~Majoris (HR~2781;
HD~57060). This is also a short-period (P=4.3934~d) eclipsing binary system.
{\it ROSAT} X-ray observations presented by
Bergh\"{o}fer \& Schmitt (1995b) showed phase-locked variability 
with a single broad trough centered near secondary minimum. 
However a colliding winds interpretation was dismissed by the authors
on the assumption that the wind momenta of the component stars were
broadly similar and thus that the shocked region was nearly planar
(Wiggs \& Gies 1993). Such a scenario should form a double-peaked
lightcurve (\eg Pittard \& Stevens 1997) and
the majority of the emission was therefore attributed to the secondary's wind.
The unremarkable X-ray luminosity 
also seemed in accordance with single O-star emission.
However, if the wind momenta are much more imbalanced than believed, 
the lightcurve {\em would} be expected to have a single broad minimum, in
agreement with the observations. HD~57060 also suffers from the Struve-Sahade
effect (Stickland 1997; Gies, Bagnuolo \& Penny 1997), and whilst the cause 
of this remains uncertain, colliding stellar winds is one possible 
interpretation. We therefore conclude that 
colliding winds may still be significant in this system.

Other strong candidates for colliding winds emission include HD~93205 and
HD~152248 (\cf Corcoran 1996), both of which are due to be observed with 
the {\it XMM} satellite, and HD~165052 (see Pittard \& Stevens 1997 
for more details).

\subsection{Those with no colliding winds signature}
Many early-type binaries observed by X-ray satellites show no
signature of colliding winds emission. For the most part this is
because the observations are generally short, the lightcurves
extremely sparse and the expected luminosities near or below previous detection
limits. However, one fairly bright frequently observed system 
which shows no signs whatsoever of colliding winds emission is 
$\delta$~Orionis~A (HD~36486), a spectroscopic and eclipsing binary (O9.5~II +
B0~III) with a 5.7~day period. As far as we know there has been
no published paper which specifically investigates any potential 
colliding winds emission in this system.

\section{Conclusions}
\label{sec:conclusions}

\begin{table*}
\caption{Summary of the strengths and weaknesses of the two interpretations
of the X-ray emission considered.}
\label{tab:conclusions}
\begin{center}
\begin{tabular}{|l|l|l|}
\hline
     & Intrinsic Wind Shocks & Colliding Winds \\
\hline
Pros & Lack of variability in {\it ASCA} data. & Lightcurve and spectra of model cw\_1 are consistent\\ 
     &                                         & with the observational data, despite neither matching\\
     &                                         & our initial simple expectations. \\
\vspace{2mm} \\
     & Low mass-loss rates of the stars can reduce CW emission & \\
     & but without running into problems with the X-ray emission & \\
     & measure exceeding that of the wind. & \\
\hline
Cons & Circumstellar attenuation around periastron may be & Modelling is very difficult and contains a number of \\
     & significant.                                       & assumptions and approximations. \\
\vspace{2mm} \\
     & & Optical data suggests that the wind collision shock \\
     & & does not crash onto the surface of the secondary \\
     & & during periastron passage. \\
\hline
\end{tabular}
\end{center}
\end{table*}

\noindent In this paper we have analysed two {\it ASCA} X-ray 
observations of the highly eccentric early-type binary 
Iota~Orionis, which were taken half an orbit apart at periastron and
apastron. Although all previous observations were short and without
good phase coverage, it was expected that a strong colliding winds
signature would be seen. This expectation was based on the almost 
order of magnitude variation in the orbital separation between 
these phases and the known dependence of the X-ray luminosity 
of colliding winds (and suspected dependence of the
spectral shape) with this. In turn it was hoped that this would
reveal further insight into the physics of radiative driving and 
stellar winds from early-type stars, and in particular the additional 
interactions possible in binary systems. Indeed, Iota~Orionis was 
selected as a target precisely because substantial changes were 
expected which would allow us to `probe' the dynamics of the wind 
collision and infer the amount of radiative braking/inhibition.

Although our analysis was complicated by various problems experienced
with the {\it ASCA} datasets, we found the emission to be surprisingly
constant and unvarying between the two observations, in direct
contradiction with our expectations. A further analysis of archival 
{\it ROSAT} data demonstrated a relatively low X-ray luminosity from
this system of $L_{x} \sim 1.0 - 1.6 \times 10^{32} \ergps$. Using a simple 
model we confirmed that the column along the line of sight to the 
expected position of the shock apex was unlikely to produce any 
significant attenuation of the shock X-rays, except perhaps around
periastron. Based on an expected $1/D$ variation in the emission this 
did not help to explain our constant count rate, and
subsequently significantly more complex hydrodynamical models were used to
investigate the problem. 
 
Despite strongly phase-variable emission from the models, both were
consistent with the observations. The model with the lower mass-loss 
rates (cw\_2) predicted attenuated luminosities an order of magnitude below 
the observational data, implying that intrinsic wind shocks emit the
majority of the observed emission. On the other hand,
the model with the higher mass-loss rates (cw\_1)
predicted minima in the colliding winds emission which coincided with
the phases of the {\it ASCA} observations, and slightly enhanced
emission at the phase of the {\it ROSAT} observation. This was also
consistent with the data, implying that the observed emission could
also be interpreted as purely colliding winds emission. In 
Table~\ref{tab:conclusions} we summarize the strengths and weaknesses of
each of these.

Unfortunately
it is impossible to distinguish between these two interpretations with
the limited dataset currently available. Additional observations
are clearly needed. New data with increased spatial resolution (to
avoid source confusion from objects such as NSV~2321), spectral
resolution and counts (to constrain spectral and luminosity variability),
and phase coverage is necessary if we are to begin distinguishing the
competing interpretations. Observations at other wavelengths to
further refine the system parameters (particularly the mass-loss rates
of {\em both} stars) are also desirable.
 
Finally we compare the Iota~Orionis observations with those 
obtained for other early-type binary systems. Although Iota~Orionis is
not the first well-studied early-type binary to show no {\em clear} 
signature of colliding winds emission ($\delta$~Orionis~A is another), 
it {\em is} the first binary with a highly eccentric orbit not to do so. 
The lack of a clear colliding winds signature in Iota~Orionis makes 
it noticeably different from other eccentric binaries (\eg WR~140,
$\gamma^{2}$~Velorum) where such signatures are readily seen at X-ray
energies. Thus, Iota~Orionis raises new open questions in our 
understanding of early-type binaries. In particular, we would like to 
know why the colliding winds emission is not obvious in Iota~Orionis. 
Is it simply a result of the low mass-loss rates, a particular
combination of system parameters, or are there deeper 
forces at work? This may be the first indication of the importance
of radiative braking on a line-driven wind.

\section*{Acknowledgments}
\label{sec:acknowledgments}
We thank the {\it XMEGA} group for the opportunity of analysing the
{\it ASCA} datasets and John Blondin for the use of VH-1. In
particular we would like to acknowledge some very helpful comments 
from various members of the {\it XMEGA} group.
We acknowledge the use of the Birmingham and Leeds {\it Starlink} 
nodes where the calculations were performed. We
are indebted to the developers and maintainers of the X-ray packages
{\sc XSPEC} and {\sc XSELECT}, distributed by HEASARC, and of 
{\sc ASTERIX}, distributed by {\sc PPARC}. JMP gratefully 
acknowledges funding from the School's of
Physics \& Astronomy at Birmingham and Leeds, and IRS acknowledges 
funding of a PPARC Advanced Fellowship. This research has made use of 
the SIMBAD astronomical database at the CDS, Strasbourg. Finally we would
like to thank the referee whose constructive comments greatly improved
this paper.

\label{lastpage}
\end{document}